\documentclass[aps,pre,epsf,superscriptaddress,amsmath,amssymb,amsfonts,twocolumn,showpacs]{revtex4-1}
\usepackage{amsmath,epsfig}
\usepackage{amssymb,amsfonts}
\usepackage{xcolor}
\usepackage{verbatim}
\usepackage{graphicx}
\usepackage{bm}
\usepackage{fouridx}
\usepackage[titletoc]{appendix}

\usepackage[T1]{fontenc} 
\usepackage{graphicx}
\usepackage{epsfig}
\usepackage{dcolumn}
\usepackage{bm}
\usepackage{braket}
\usepackage{amsmath}
\usepackage{mathtools}
\usepackage{color}
\usepackage{hyperref}
\newcommand{\abs}[1]{\left| #1 \right|} 

\usepackage{hyperref}
\usepackage{multirow}

\begin{document}
\title{Many-body effects on second-order phase transitions in\\ spinor Bose-Einstein condensates and breathing dynamics}

\author{K. M. Mittal}
\affiliation{Indian Institute of Science Education and Research,
Dr. Homi Bhabha Road, Pune 411 008, India} 
\affiliation{Center for Optical Quantum Technologies, Department of Physics, University of Hamburg, 
Luruper Chaussee 149, 22761 Hamburg Germany}
\author{S. I. Mistakidis}
\affiliation{Center for Optical Quantum Technologies, Department of Physics, University of Hamburg, 
Luruper Chaussee 149, 22761 Hamburg Germany}
\author{P. G. Kevrekidis}
\affiliation{Department of Mathematics and Statistics, University of Massachusetts, Amherst, 
Massachusetts 01003-4515, USA}
\affiliation{Mathematical Institute, University of Oxford, Oxford, OX2
  6GG, UK}
\author{P. Schmelcher}
\affiliation{Center for Optical Quantum Technologies, Department of Physics, University of Hamburg, 
Luruper Chaussee 149, 22761 Hamburg Germany}
\affiliation{The Hamburg Centre for Ultrafast Imaging,
University of Hamburg, Luruper Chaussee 149, 22761 Hamburg, Germany}

\date{\today}

\begin{abstract} 

We unravel the correlation effects of the second-order quantum phase transitions emerging on the ground state of a harmonically trapped spin-1 Bose gas, upon varying the involved Zeeman terms, as well as its breathing dynamics triggered by quenching the trapping frequency. 
It is found that the boundaries of the associated magnetic phases are altered in the presence of interparticle correlations for both ferromagnetic and anti-ferromagnetic spin-spin interactions, an effect which becomes more prominent in the few-body scenario. 
Most importantly, we unveil a correlation-induced shrinking of the anti-ferromagnetic and broken-axisymmetry phases implying that ground states with bosons polarized in a single spin-component are favored. 
Turning to the dynamical response of the spinor gas it is shown that its breathing frequency is independent of the system parameters while correlations lead to the formation of filamentary patterns in the one-body density of the participating components. 
The number of filaments is larger for increasing spin-independent interaction strengths or for smaller particle numbers. 
Each filament maintains its coherence and exhibits an anti-correlated behavior while distinct filaments show significant losses of coherence and are two-body correlated. 
Interestingly, we demonstrate that for an initial broken-axisymmetry phase an enhanced spin-flip dynamics takes place which can be tuned either via the linear Zeeman term or the quench amplitude. 
\end{abstract}

\maketitle

\section{Introduction} 

Since the first realization of an optically trapped Bose-Einstein condensate (BEC) of ${}^{23}$Na atoms with spin degrees-of-freedom \cite{and_98}, spinor BECs have been among the most actively studied
systems within the ultracold realm \cite{ho1998spinor,intro_1,kawaguchi2012spinor,stamper2013spinor}. 
The rich phase diagram emerging in these setups \cite{stenger1998spin,phase_2_2,phase_2,phase_3} renders them
particularly important for understanding quantum phase transitions \cite{bookjans2011quantum,chudnovskiy2018symmetry}. 
Indeed, various studies have been devoted to exploring the properties of the associated magnetic phases \cite{zhang2003mean,spin1_phase,jacob2012phase}. 
Of interest has also been the examination of the boundaries between metastable spin domains in a spin-1 Bose gas via measuring the involved tunneling rates \cite{tun_1}, the investigation of the phase diagram for a spin-2 Bose gas using spin transfer processes \cite{2_phase}, as well as the induction of phase separation phenomena in spin-3 Bose gases \cite{3_spin}.
Such spin systems have also been central in the exploration of
topological excitations in the form of
skyrmions and merons~\cite{bigelow}, monopoles~\cite{dshall} and knots~\cite{motto2}, among others.

Another interesting consequence of the inclusion of internal degrees-of-freedom is the presence of spin-mixing dynamics in these systems due to
spin-exchange collisions \cite{kronjager2005evolution,cui2008quantum,barnett2010antiferromagnetic,chang2005coherent,black2007spinor,gu2007coherent,yukalov2018dipolar}. 
This process gives rise to coherent and reversible transfer of atoms between
the magnetic sublevels of the system while its total spin is conserved. 
Such spin dynamics has been observed in spin-1 ${}^{87}$Rb \cite{chang2004observation} and ${}^{23}$Na \cite{stenger1998spin}, spin-2 ${}^{87}$Rb \cite{spin-col_2,spin-col_2_2} as well as spin-3 ${}^{52}$Cr atomic gases \cite{spin-col_3}. 
Notable implementations of the emergent spin dynamics range from interferometry applications \cite{int_2,int_1}, 
entanglement generation \cite{ent_2,xu2019efficient}, formation of spin domains and spin textures \cite{dom_2,dom_3,dom_1,guzman2011long} to the
realization of soliton complexes \cite{dom_3,dom_1,sol_4,sol_3,sol_5}. 
Most importantly, the presence of multiple magnetic phases and spin-mixing dynamics makes these systems a particularly 
interesting playground for studying out-of-equilibrium phenomena induced e.g. by quantum quenches
\cite{sadler2006spontaneous,leslie2009amplification,bookjans2011quantum,quench_1,kang2020crossover}.

A unifying aspect of most of the above-mentioned studies is their reliance on the mean-field (MF) approximation, resting under the premise of a macroscopic wavefunction for each component. 
Despite the success of this framework in describing several phenomena \cite{ho1998spinor,ohmi1998bose,stenger1998spin}, an additional more recent
focus has been on understanding the effect of correlations emerging in these systems \cite{cui2008quantum,leslie2009amplification,konig2018quantum,saha2020strongly}. 
Pioneering works include the study of universality in the spin-dynamics of spinor ${}^{87}$Rb BECs \cite{prufer2018observation,uni_2,prufer2019experimental} where information regarding the presence of correlations emanating in these systems has been experimentally obtained. 

Motivated by these experimental efforts, here we unravel correlation driven phenomena in the ground state properties and the quench dynamics of one-dimensional (1D) harmonically trapped spin-1 Bose gases. 
To achieve this we systematically compare the underlying ground state magnetic phases between the MF approximation, where interparticle correlations are neglected, and a many-body (MB) variational treatment. 
Another pivotal point that we touch upon within our work is how the size of a confined system
affects the transition boundaries between the different magnetic phases. 
Moreover, we unravel the imprint of correlations \cite{cui2008quantum,leslie2009amplification} in the breathing dynamics of the spin-1 Bose gas 
following a quench of the external harmonic oscillator frequency. 
To track the ground state properties and the correlated quantum dynamics of the bosonic spinor gas we utilize the variational Multi-Layer Multi-Configuration Time-Dependent Hartree method for multicomponent systems (ML-MCTDHX)\cite{cao2017unified,bolsinger2017beyond,cao2013multi}. 
The latter enables us to address all the relevant interparticle correlations of the spinorial system.

Regarding the ground state phase diagram of the spin-1 Bose gas we find that there is no noticeable
change in the involved first-order transitions, while the boundaries of the second-order ones are 
significantly altered in the presence of correlations. 
In other words, interparticle correlations are non-neglible only for second-order phase transitions 
where a superposition of spin-states contributes to the ground state of the system. 
In particular, the boundaries of the second-order transitions are considerably shifted for both 
ferromagnetic \cite{chang2004observation} and anti-ferromagnetic \cite{inouye1998observation} spin-spin interactions. 
This is a phenomenon that crucially depends on the finite size of the system  since correlation-induced phenomena 
become more pronounced in the few-body case. 
Remarkably a shrinking of the antiferromagnetic and broken-axisymmetry phases, which is explicitly driven by the interparticle correlations, occurs favoring this way ground states with bosons being polarized in a single spin-component. 
Furthermore we recover the theoretical MF predictions for an adequately large number of bosons, thus further revealing the crucial role of interparticle correlations 
in setups containing a finite particle number \cite{cui2008quantum,koutentakis2019probing}. 

Turning to the breathing dynamics \cite{breathing_1,pyzh2018spectral,mistakidis2018correlation} 
of the spinor gas, following a quench of its trapping frequency, we unveil that it exhibits almost the same frequency for all participating components both within the MF and the MB approach as well as for different initial phases. 
In sharp contrast to the MF approximation, it is shown that the inclusion of correlations leads to 
the formation of filamentary-like patterns \cite{mistakidis2019quench,mistakidis2018correlation,erdmann2019phase} 
in the density profile of each component. 
These refer here to a multihump structure building upon the background 
density of the Bose gas and become more prominent for smaller particle numbers or for 
increasing spin-independent interaction strengths. 
Importantly, we showcase that each filament corresponds to a coherent structure while 
for neighboring filaments significant losses of coherence occur. 
Moreover, we find that two particles within the same filament are anti-correlated whereas particles 
residing in neighboring filaments are correlated with one another. 
Interestingly, for quenches within the broken-axisymmetry phase, where all spin-components are populated, 
spin fluctuations \cite{spin_fluctuations} are manifested implying a transfer of particles between the 
individual components. 
We demonstrate that this intercomponent trasmission process can be controlled by tuning either the value of the linear Zeeman term or the quench amplitude. 
This modification of the spin-component populations
is a feature enabled in this setting that is naturally absent in both single
and two-component condensates (in the latter in the absence of Rabi coupling).

Our presentation is structured as follows. 
Section~\ref{Setup} introduces the relevant theoretical framework and spin operators while 
section~\ref{comp_meth} describes the employed MB methodology and its ingredients. 
In section~\ref{phase_space} we discuss in detail the correlation effects in the ground state phase diagram 
of the spin-1 Bose gas. 
Section~\ref{quench} presents the correlated breathing dynamics of the spinor gas after quenching 
the trap frequency to lower values. 
Finally, in section~\ref{conclusion} we summarize our results and provide an outlook onto future perspectives.

\section{Setup and spin operators}
\label{Setup}

We consider an ultracold spinor $F=1$ (alias spin-1) Bose gas consisting of $N$ bosons with mass $M$ and 
being trapped in a 1D harmonic trap. 
We aim to investigate the underlying ground state phase diagram of this system and its emergent out-of-equilibrium breathing dynamics when interparticle correlations are taken into account and the beyond MF contributions may become significant. 
The MB Hamiltonian of such a system can be written as the sum $\hat{H}=\hat{H}_0+\hat{V}$, where its non-interacting part $\hat{H}_0$ reads
\begin{equation} 
\begin{split}
    \hat{H}_0= \int dx \sum_{\alpha,\beta=-1}^{1} &\hat{\psi}_{\alpha} ^ {\dagger}(x) \Big[ -\frac{\hbar^2}{2M}\frac{\partial^2}{\partial x^2} + \frac{1}{2}M\omega^2x^2\\&-p(f_z)_{\alpha\beta}+q(f_z^2)_{\alpha\beta}\Big] \hat{\psi}_{\beta}(x). \label{noninteracting_part}
\end{split}
\end{equation}
Here $\hat{\psi}_\alpha(x)$ denotes the bosonic field operator accounting for the magnetic sublevels with spin-$z$ projection $\alpha= \{-1,0,1 \}$ (alias components) of the $F=1$ hyperfine manifold. 
Also, $\omega=0.1$ is the trapping frequency and $(f_z)_{\alpha\beta}=\alpha\delta_{\alpha\beta}$ are the matrix elements of the spin-$z$ Pauli matrix. The quantity $\omega$ here, as is customary in
one-dimensional system reductions~\cite{pethick2008bose,pitaevskii2016bose},
plays the role of the longitudinal over the transverse trapping frequencies
and is typically $\omega \ll 1$. 
Additionally $p$, $q$ refer to the corresponding linear and quadratic Zeeman energy shift parameters respectively. 
They can be experimentally tuned by either adjusting the applied magnetic field \cite{p_q_tune_1} or using a microwave dressing field \cite{leslie2009amplification,bookjans2011quantum}. 
In particular, the linear (quadratic) Zeeman energy term is linearly (quadratically) proportional to the external magnetic field applied along the $z$-direction, see e.g. Refs. \cite{kawaguchi2012spinor,zhang2003mean} for more details. 
These terms essentially lead to an effective detuning of the $\alpha=\pm 1$ components with respect to the $\alpha=0$ one. 

The interacting part, $\hat{V}$, of the Hamiltonian~\cite{yukalov2018dipolar} is given by
\begin{equation}
    \hat{V}= \frac{1}{2}\int dx \Big[ c_0 :\hat{n}^2(x):+ c_1 :\hat{F}^2(x):\Big],\label{interaction_part}
\end{equation}
where the symbol : : denotes the well-known normal ordering of the involved operators which essentially leads to the annihilation operators being placed to the right-hand-side of the creation ones \cite{sakurai1967advanced,ryder1996quantum,brown1994quantum}. 
The interaction strength parameters are expressed as $c_0=\frac{4\pi\hbar^2(a_0+2a_2)}{3Ma_\perp}$ and $c_1=\frac{4\pi\hbar^2(a_2-a_0)}{3Ma_\perp}$, where $a_0$ and $a_2$ refer to the three-dimensional $s$-wave scattering lengths of the atoms in the scattering channels characterized by total spin $F=0$ and $F=2$ respectively \cite{romero2019controlled}, see also Eq. (\ref{total_spin_operator}) below. 
Recall that since we operate in the ultracold regime $s$-wave scattering constitutes the dominant interaction process. 
More specifically, $c_0$ is the spin-independent interaction strength whose positive (negative) values account for repulsive (attractive) interparticle interactions. 
In contrast, $c_1$ corresponds to the spin-dependent interaction strength which is positive (negative) for anti-ferromagnetic (ferromagnetic) interactions \cite{ho1998spinor}, see also the discussion below. 
In a corresponding experiment, $c_1$ can be adjusted by using the microwave-induced Feshbach resonance technique~\cite{fesh_1}. Additionally, $a_\perp=\sqrt{\hbar/(M\omega_\perp)}$ is the transversal confinement length scale and $\omega_\perp$ the corresponding trapping frequency. 
The latter can be experimentally tuned with the aid of confinement induced resonances \cite{res_1,res_2}. 

Moreover, the total spin operator reads
\begin{equation}
\begin{split}
  \hat{F}^2=\int dx\int dy & \sum_{\alpha,\beta,\gamma,\delta=-1}^{1} \sum_{i \in \{x,y,z\}}^{}(f_i)_{\alpha\beta}(f_i)_{\gamma\delta}
  \\& \times \hat{\psi}^{\dagger}_\alpha (x)\hat{\psi}_\beta (x)\hat{\psi}^{\dagger}_\gamma (y)\hat{\psi}_\delta (y). \label{total_spin_operator}
\end{split}
\end{equation}
The expectation value of this operator provides the total spin of the system. 
The square of the  normal ordered spin density operator $\hat{F}(x)$ has the form
\begin{equation}
\begin{split}
  :\hat{F}^2(x):=\sum_{\alpha,\beta,\gamma,\delta=-1}^{1} &\sum_{i \in \{x,y,z\}}^{}(f_i)_{\alpha\beta}(f_i)_{\gamma\delta} 
\\& \times  \hat{\psi}^{\dagger}_\alpha (x)\hat{\psi}^{\dagger}_\gamma (x)\hat{\psi}_\delta (x)\hat{\psi}_\beta (x). 
  \end{split}
\end{equation}
In these expressions, the Pauli-$x$ and $y$ matrix elements are $(f_x)_{\alpha\beta}=\delta_{\alpha,\beta+1}+\delta_{\alpha,\beta-1}$ and $(f_y)_{\gamma\delta}=-i\delta_{\gamma,\delta+1}+i\delta_{\gamma,\delta-1}$ respectively. 
The indices $\alpha,\beta,\gamma,\delta \in \{ -1,0,1 \}$ refer to the individual spin components along a particular ($x$, $y$, $z$) spin direction. 
The expectation value of this operator refers to the diagonal of the spatially resolved spin-spin correlator of the system. 
On the other hand, the square of the normal ordered particle density operator $\hat{n}(x)$ is
\begin{equation}
  :\hat{n}^2(x):=\sum_{\alpha,\beta=-1}^{1}\hat{\psi}^{\dagger}_\alpha (x)\hat{\psi}^{\dagger}_\beta (x)\hat{\psi}_\beta (x)\hat{\psi}_\alpha (x),
\end{equation} 
with the summation being performed over all spin $\alpha,~\beta \in \{ -1,0,1 \}$ components. 
The expectation value of this operator corresponds to the diagonal of the spatially resolved two-particle density of the spinor system integrating out all three ($\alpha=-1,0,1$) spin components. 

In the following, the MB Hamiltonian of the spinor system is rescaled in units of $\hbar\omega_\perp$. 
Consequently, the corresponding length, time and interaction strengths are expressed in terms of $\sqrt{\hbar/ (M\omega_\perp)}$, $\omega_\perp^{-1}$ and $\sqrt{ \hbar^3\omega_\perp/M}$ respectively. 
Importantly, the experimentally relevant values of $c_1= 0.018\sqrt{\hbar^3\omega_\perp/M}$, $c_0=0.5\sqrt{\hbar^3\omega_\perp/M}$ corresponding to the spin-dependent and spin-independent interaction strengths between the atoms of ${}^{23}$Na, is taken for exploring an anti-ferromagnetic ($c_1>0$) condensate \cite{schmied2020stability,dom_1}.
On the other hand, for the ferromagnetic ($c_1<0$) case we use $c_1= -0.0047\sqrt{\hbar^3\omega_\perp/M}$ and $c_0=1\sqrt{\hbar^3\omega_\perp/M}$ which correspond to the spin-dependent and spin-independent coupling constants respectively between ${}^{87}$Rb atoms \cite{schmied2020stability,romero2019controlled}.

\section{Many-body Wavefunction Ansatz and reduction to the mean-field approximation}
\label{comp_meth} 

Our approach to calculate the ground state properties as well as to monitor the nonequilibrium quantum dynamics of the 
spinor Bose gas relies on the ML-MCTDHX method \cite{cao2017unified,bolsinger2017beyond,cao2013multi}. 
It is an {\it{ab-initio}} variational method \cite{lode2019multiconfigurational} for solving the time-dependent MB Schr\"{o}dinger equation, $\left( {i\hbar {\partial _t} - \hat H}\right) |\Psi (t) \rangle= 0$, of multicomponent systems with either bosonic \cite{mistakidis2018correlation,katsimiga2017dark,mistakidis2020induced} or fermionic \cite{kwasniok2020correlated,mistakidis2019repulsive,erdmann2019phase} constituents possessing also spin degrees-of-freedom \cite{mistakidis2019quench,mistakidis2020many,koutentakis2019probing}. 
The advantage of ML-MCTDHX is the expansion of the MB wavefunction with respect to a time-dependent and variationally optimized MB basis set which in turn allows for the optimal truncation of the relevant Hilbert space of the system. 
Accordingly, its ansatz is tailored to capture all the important intra- and intercomponent correlations of systems with mesoscopic particle numbers in a computationally efficient manner. 

More specifically the MB wavefunction ansatz, $| \Psi (t) \rangle$, is expressed as a linear combination of time-dependent permanents, $|\vec{n} (t) \rangle$, with time-dependent weight coefficients $A_{\vec{n}}(t)$. 
Namely it reads  
\begin{equation}
    | \Psi (t) \rangle =\sum_{\vec{n}} A_{\vec{n}}(t) | \vec{n} (t) \rangle.
    \label{eq:mb_wfn}
  \end{equation}
Each time-dependent permanent, with occupation numbers $\vec{n}=(n_1,\dots,n_D)$, is expanded in terms of $D$ time-dependent variationally optimized single-particle spin-orbitals (SPSOs) i.e. $\Phi_j(x,\alpha;t)$, where $\alpha =  +1,0,-1 $ and $j=1,2,\dots,D$. 
This expansion allows us to capture the interparticle correlations within and among the individual spin components.  

Subsequently, the SPSOs are expressed in a basis spanned by $d$ distinct time-dependent single-particle functions (SPFs) $\lbrace \phi_k(x;t) \rbrace$. 
The latter possess information only on the spatial state of the particle and are independent from the three-dimensional spin basis i.e. $\{ | +1 \rangle, |0\rangle, | -1 \rangle \}$ for the $F=1$ degree-of-freedom. 
Therefore, the SPSOs are given by 
\begin{equation}
    \Phi_j(x,\alpha;t)= \sum_{k=1}^{d} B^j_{k \alpha}(t) \phi_k(x;t),
    \label{eq:spso}
\end{equation}
where $B^j_{k \alpha}(t)$ refer to the corresponding time-dependent expansion coefficients. 
In this way, the correlations between the spin and spatial degrees-of-freedom are taken into account. 
Moreover, each $\phi_k(x;t)$ is expressed in terms of a discrete variable representation (DVR). 
The time-evolution of the $N$-body spinor wavefunction governed by the MB Hamiltonian $\hat{H}$ [Eqs. (\ref{noninteracting_part}) and (\ref{interaction_part})] reduces to the determination of the $A$-vector coefficients, the $B^j_{k \alpha}(t)$ expansion coefficients of the SPSOs and the SPFs $\phi_k(x;t)$. 
These in turn follow the variationally obtained ML-MCTDHX equations of motion, see for details      \cite{cao2017unified,bolsinger2017beyond}. 
The latter consist of a set of $\binom{N+D-1}{D-1}$ ordinary linear differential equations for the $A$-vector coefficients, coupled to $D$ and $d$ non-linear integrodifferential equations for the SPSOs and SPFs respectively. 

Another notable feature of ML-MCTDHX is that it enables us to operate within different correlation levels. 
As a case example, in the limiting case of $D=1$ and $d=3$ accounting for the hybridization of the spin and spatial degrees-of-freedom it reduces to the time-dependent Gross-Pitaevskii equation for a three-component spinor system \cite{schmied2020stability,pethick2008bose,pitaevskii2016bose,sol_5}. 
Indeed within this limit only a single SPSO is involved and therefore the MB ansatz boils down to the MF product state, namely 
$\Psi(x_1,\alpha_1,x_2,\alpha_2,...,x_N,\alpha_N;t)=\prod_{i=1}^{N} \Phi_1(x_i,\alpha_i,t)$. 
Note that ($x_1$, $x_2$, $\dots$, $x_N$) refer to the spatial coordinates of the particles characterized by the corresponding spin configuration ($\alpha_1$, $\alpha_2$, $\dots$, $\alpha_N$). 
Employing a variational principle~\cite{yukalov2018dipolar} for this latter MF ansatz we can easily retrieve the well-known coupled system of Gross-Pitaevskii equations of motion for the different hyperfine states \cite{romero2019controlled,schmied2020stability} described by the individual spin orbitals $\Phi_1(x,\alpha;t)$ with $\alpha=\pm1,0$ which are independent spatial functions. 
In particular, the $\alpha=+1$ and the $\alpha=-1$ components obey   
\begin{widetext}
\begin{equation}
\begin{split}
    i\partial_t\Phi_1(x,\pm1;t)=\Big(-\frac{1}{2}\frac{\partial^2}{\partial x^2}+\frac{1}{2}\omega^2 x^2 \mp p+ q\Big)\Phi_1(x,\pm1;t)+c_0\sum_{\alpha=-1} ^{1}|\Phi_1(x,\alpha;t)|^2\Phi_1(x,\pm1;t) \\+c_1(|\Phi_1(x,\pm1;t)|^2+|\Phi_1(x,0;t)|^2-|\Phi_1(x,\mp1;t)|^2)\Phi_1(x,\pm1;t) +c_1\Phi_1^2(x,0;t)\Phi_1^{*}(x,\mp1;t),
\end{split}
\end{equation}
while the $\alpha=0$ spin-state satisfies 
\begin{equation}
\begin{split}
    i\partial_t\Phi_1(x,0;t)=\Big(-\frac{1}{2}\frac{\partial^2}{\partial x^2}+\frac{1}{2}\omega^2 x^2\Big)\Phi_1(x,0;t)+c_0\sum_{\alpha=-1} ^{1}|\Phi_1(x,\alpha;t)|^2\Phi_1(x,0;t) \\+c_1(|\Phi_1(x,+1;t)|^2+|\Phi_1(x,-1;t)|^2)\Phi_1(x,0;t) + 2c_1\Phi_1(x,+1;t)\Phi_1^{*}(x,0;t)\Phi_1(x,-1;t).
\end{split}
\end{equation}
\end{widetext}
On the other hand, in the case of $D=3 M_{p}$, $d=M_{p}$, where $M_p$ is the dimension of the DVR basis, the 
ML-MCTDHX method is equivalent to a full configuration interaction approach, commonly referred to in the literature as ``exact diagonalization''.

For our implementation we have used a sine DVR as a primitive basis for the SPFs including $M_p=600$ grid points. 
Note that the sine-DVR inherently introduces hard-wall boundary conditions at its endpoints. 
In particular, we have employed hard-wall boundaries at positions $x_{\pm}=\pm50$ for $N=50$ particles, $x_{\pm}=\pm35 $ for $N = 20$ and $x_{\pm}=\pm25 $ for $N=5$. 
Of course, we have ensured while choosing the location of these boundaries that they do not affect our results since there are not appreciable densities e.g. beyond $x_{\pm}=\pm20$ for $N=50$. 
To study the dynamics, we propagate the MB wavefunction by utilizing the appropriate Hamiltonian within the ML-MCTDHX equations of motion. 
The accuracy of the results obtained within the ML-MCTDHX approach has been confirmed by verifying that the observables of interest become almost insensitive (within a given level of accuracy) upon varying the number of used SPSOs $D$ and SPFs $d$. 
More specifically, in the following we employ $D=6$, $d=6$ for all cases. 
For instance, comparing the one-body density of each component $\rho^{(1)}_{\alpha}(x;t)$ for $p/(|c_1|n)=0.04$, $q/(|c_1|n)=-0.44$, $c_0=1\sqrt{\hbar^3\omega_\perp/M}$ depicted in Fig. \ref{fig:quench_5} between the ($D=6$, $d=6$) and the ($D=8$, $d=8$) cases we have found that the corresponding relative deviation lies below $4\%$ throughout the evolution. 
\begin{figure*}[t]
       \includegraphics[width=0.8\textwidth,height=\textheight,keepaspectratio]{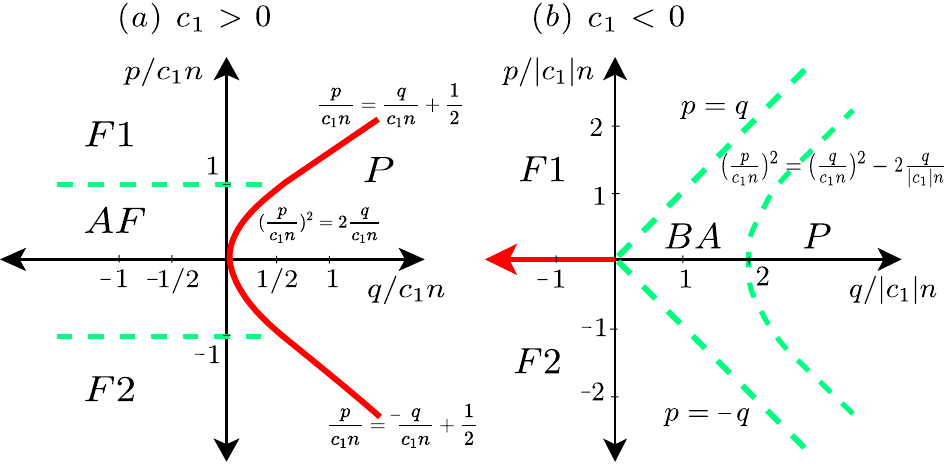}
        \caption{Schematic representation of the ground state phase diagram of the spin-1 Bose gas for (a) anti-ferromagnetic $c_1>0$ and (b) ferromagnetic $c_1<0$ spin-dependent interactions and varying linear $p$ and quadratic $q$ Zeeman energy shift parameters in the thermodynamic limit. 
        Recall that in this latter case the hybridization of the spin and the spatial degrees-of-freedom is neglected. 
        Red solid lines denote the boundaries of the emergent first-order quantum phase transitions, while the green dashed lines show the boundaries of the second-order quantum phase transitions. 
        The symbol $n$ refers to the total density of the Bose gas. 
        Depending on the sign of $c_1$ the phase diagram involves two ferromagnetic phases (F1), (F2), an anti-ferromagnetic (AF), a polar (P) and a broken axisymmetry (BA) phase.
      } 
        \label{fig:phase_dia}
\end{figure*}

\section{Ground state phase diagram of the spin-1 Bose gas}
\label{phase_space} 

It is known that within the thermodynamic limit $N\to \infty$ and in the MF realm, where the interparticle correlations of the spin-1 Bose gas are ignored, the interplay between the sign of the spin-dependent interaction $c_1$ and the strength of the Zeeman energy terms $p$, $q$ [see also Eq. (\ref{noninteracting_part})] results in a rich ground state phase diagram \cite{schmied2020stability,kawaguchi2012spinor,jacob2012phase,stamper2013spinor}. 
A schematic representation of the aforementioned phase diagram, the involved phases and the underlying phase transition boundaries when $N\to \infty$ are depicted in Fig. \ref{fig:phase_dia}. 
Indeed for anti-ferromagnetic interactions $c_1>0$, there are two ferromagnetic phases with the particles residing either in the $\alpha=+1$ (F1) or the $\alpha=-1$ (F2) component. 
Moreover, there is an anti-ferromagnetic phase (AF), in which the particles populate both the $\alpha=+1$ and the $\alpha=-1$ states in a non-equal fashion while the $\alpha=0$ component remains completely unoccupied. 
Also, there is the Polar phase (P) where the particles are entirely in the $\alpha=0$ state. 
On the other hand, for ferromagnetic interactions namely $c_1<0$ one additional phase emerges, the so-called broken-axisymmetry phase (BA) where all the three spin-states $\alpha=\pm1, 0$ are occupied.

Importantly, by inspecting Fig. \ref{fig:phase_dia} it becomes evident that for this system there is a multitude of possible quantum phase transitions for fixed $c_1$ and varying $p$, $q$. 
Tuning the latter parameters enables us to transit from one phase to the other. 
These quantum phase transitions between the different magnetic phases, can be classified according to their continuous (second-order) or non-continuous (first-order) character \cite{carr2010understanding,sachdev1999quantum}. 
The first-order transitions are characterized by the abrupt change of the spin-state (component) $\alpha$ that contributes to the ground state of the system as the transition point is crossed. 
For instance, within the F1 phase all of the particles occupy the spin-state with $\alpha=1$ while in the P phase all atoms populate the spin-state with $\alpha=0$. 
Namely when crossing the curve of the F1 to the P phase transition [see the red line in Fig. \ref{fig:phase_dia}(a)] for a larger $q$ the spin-state contributing to the ground state of the system changes without accessing a superposition spin-state of the $\alpha=0$ and the $\alpha=1$ components. 
In sharp contrast, for second-order phase transitions the system's ground state transits from a state characterized by the occupation of a single spin-state to a superposition one where a second spin-state acquires finite population across the underlying phase boundary. 
As an example, along the transition curve from the F2 to the AF phase [see the green dashed lines in Fig. \ref{fig:phase_dia} (a)] i.e. by increasing $p$, the system initially (F2 phase) occupies the $\alpha=-1$ component and subsequently (AF phase) resides in a superposition of the $\alpha=-1$ and the $\alpha=1$ components, with a progressively increasing $\alpha=1$ component. 

Below, we compare the MF and MB ground state of the spin-1 Bose gas when taking intra- and intercomponent correlations into account and explore the effect of the finite size of the system on the respective phase diagram. 
As we shall argue, significant correlation-induced phenomena are manifested in the magnetic phases across second-order quantum phase transitions. 
We remark that the impact of correlations has also been investigated through the involved first-order transitions of the spin-1 Bose gas and it has been found that they only negligibly affect the corresponding phase boundaries, i.e. the effects of interparticle correlations are supressed. Therefore, in the following, based on the known phase diagram of the spin-1 Bose gas for $N\to \infty$, see Fig. \ref{fig:phase_dia}, we examine the impact of correlations \cite{konig2018quantum} across each of the emerging second-order phase transitions occurring for anti-ferromagnetic ($c_1>0$) [Sec. \ref{antiferro_int}] and ferromagnetic ($c_1<0$) [Sec. \ref{ferro_int}] spin-spin interactions. 
Since the participating phases are characterized by specific constraints in the population of each component we employ as explicit measures for their presence the expectation values of the polar [see Eq. (\ref{polar_operator}) below] and spin-$z$ [see Eq. (\ref{spin_oparator}) below] operators. 
These observables essentially quantify different population imbalances between the individual components and thus their combination allows us to infer the existence of each magnetic phase upon tuning the linear or the quadratic Zeeman fields, see for details below. 

\begin{figure*}[ht]
       \includegraphics[width=1.0\textwidth,height=\textheight,keepaspectratio]{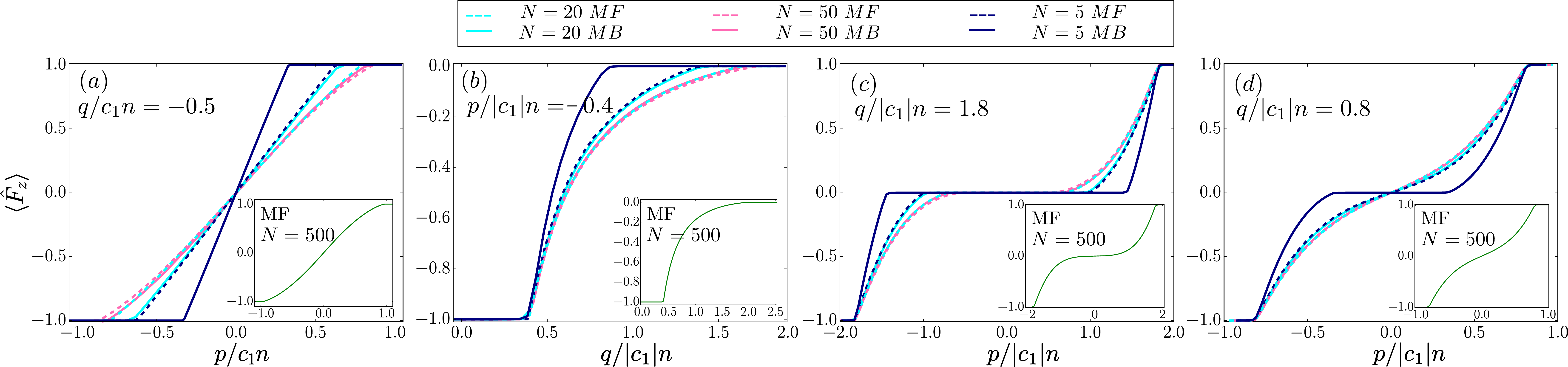}
        \caption{Population imbalance, $\langle \hat{F}_z \rangle$, between the $\alpha=1$ and the $\alpha=-1$ components showing the second-order phase transitions in the ground state of the spin-1 Bose gas. 
        $\langle \hat{F}_z \rangle$ is provided both within the MF and the MB approach as well as for distinct number of particles $N$ (see legends). 
        (a) $\langle \hat{F}_z \rangle$ illustrating the transitions between the F2, AF and the F1 phases for varying $p/(c_1 n)$ and constant $q/(c_1 n)=-0.5$ with anti-ferromagnetic interactions $c_1=0.018\sqrt{\hbar^3\omega_\perp/M}>0$ while $c_0=0.5\sqrt{\hbar^3\omega_\perp/M}$.  
        $\langle \hat{F}_z \rangle$ for ferromagnetic interactions ($c_1=-0.0047\sqrt{\hbar^3\omega_\perp/M}<0$) presenting the transitions among (b) the F2, BA and P phases for increasing $q/(|c_1| n)$ and constant $p/(|c_1| n)=-0.4$, (c) the F2, BA, P, BA and F1 phases with $q/(|c_1| n)=1.8$ and (d) the F2, BA and F1 phases with $q/(|c_1| n)=0.8$ for varying $p/(|c_1| n)$. 
        In panels (b)-(d) $c_0=1\sqrt{\hbar^3\omega_\perp/M}$. 
        The insets showcase $\langle \hat{F}_z \rangle$ in the MF method for $N=500$ bosons for the respective parameter values of the main figures.} 
        \label{fig:spin}
\end{figure*}

\subsection{Anti-ferromagnetic ensembles} \label{antiferro_int}

\begin{figure*}[ht]
       \includegraphics[width=1.0\textwidth,height=\textheight,keepaspectratio]{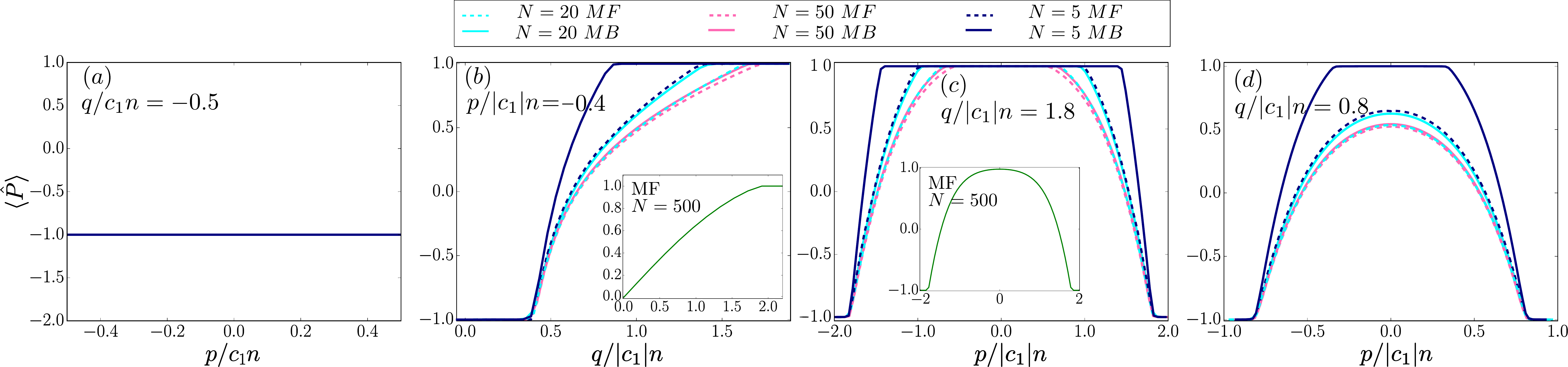}
        \caption{Population imbalance, $\langle \hat{P} \rangle$, between the $\alpha=0$ and the $\alpha= \pm 1$ components for the second-order phase transitions in the ground state of a spin-1 Bose gas both within and beyond the MF approximation and for different number of particles $N$ (see legend). 
        (a) $\langle \hat{P} \rangle$ for the transition among the F1, AF and F2 phases for varying $p/(c_1 n)$ and constant $q/(c_1 n)=-0.5$, $c_1=0.018\sqrt{\hbar^3\omega_\perp/M}>0$, $c_0=0.5\sqrt{\hbar^3\omega_\perp/M}$. 
        $\langle \hat{P} \rangle$ for ferromagnetic interactions ($c_1=-0.0047\sqrt{\hbar^3\omega_\perp/M}<0$) and $c_0=1\sqrt{\hbar^3\omega_\perp/M}$ demonstrating the transitions between (b) the F2, BA and P phases for increasing $q/(|c_1| n)$ and fixed $p/(|c_1| n)=0.4$, (c) the F2, BA, P, BA and F1 phases and (d) the F2, BA and F1 phases for different $p/|(c_1| n)$ and (c) $q/(|c_1| n)=1.8$, (d) $q/(|c_1| n)=0.8$. 
        The insets provide $\langle \hat{P} \rangle$ in the MF approach for $N=500$ bosons with respect to (b) $q/(|c_1|n)$ with fixed $p/(|c_1|n)=0$, (c) $p/(|c_1|n)$ with constant $q/(|c_1|n)=1.8$. } 
        \label{fig:polar}
\end{figure*}

In particular for anti-ferromagnetic interactions i.e. $c_1>0$, we focus on the second-order quantum phase transitions which are known to occur in the $q<0$ region as well as for some specific regions of $q/(c_1n)>0$, see remark \footnote{In particular the corresponding phases featuring a second-order phase transition in the interval $q/(c_1n)>0$ occur in the following regions. The ferromagnetic phase F1 exists for $p/(c_1n) > 1$, $0<q/(c_1n)<1/2$ as well as for $p/(c_1n) > 1$, $q/(c_1n)>1/2$ and $p/(c_1n)-q/(c_1n)-1/2>0$. 
Along the same lines, the F2 appears for $p/(c_1n) < -1$, $0<q/(c_1n)<1/2$ and for $p/(c_1n) < -1$, $q/(c_1n)>1/2$ and $p/(c_1n) + q/(c_1n) -1/2<0$. 
Finally the AF phase takes place when $-1<p/(c_1n)<1$, $0<q/(c_1n)<1/2$ and $(p/(c_1n))^2 - 2q/(c_1n)>0$ is satisfied.} for further details, involving the phases F1, F2 and AF, see Fig. \ref{fig:phase_dia} (a). 
In the following, we consider $q/(c_1n)=-0.5$ as a representative value of the quadratic Zeeman term in order to realize the above-mentioned phase transitions. 
However, we have checked that also for other values of $q/(c_1n)$ the boundaries of the phases, to be presented below, do not alter e.g. for $q/(c_1n)=-0.1$ and $q/(c_1n)=-1$.
As already explained above, for all of these phases we expect the $\alpha=0$ component to be unoccupied. 
To explicitly demonstrate that the spin-state with $\alpha=0$ is not populated throughout the above-mentioned transitions we invoke the expectation value of the polar operator which reads 
\begin{equation}
\begin{split}
 \braket{\hat{P}(t)}=&\braket{\Psi(t) |\sum_{\alpha \beta}^{} \int dx \hat{\psi}_{\alpha}^\dagger(x) P^{0}_{\alpha \beta} \hat{\psi}_{\beta}(x) | \Psi(t) } \\& = \braket{\hat{n}_0(t)} - (\braket{\hat{n}_{1}(t)}+\braket{\hat{n}_{-1}(t)}). \label{polar_operator} 
\end{split}
\end{equation}
Here, $P^0_{\alpha\beta}=(1-2|\alpha|)\delta_{\alpha\beta}$, with $\alpha,\beta \in \{-1,0,1\}$ indexing the spin components along the spin-$z$ direction and $\hat{n}_{\alpha}$ corresponds to the particle number operator of the $\alpha^{th}$ spin-state. 
This expectation value, $\langle\hat{P}(t)\rangle$, quantifies the population difference between the number of atoms in the $\alpha=0$ spin-state to that of the ones residing in the $\alpha=\pm1$ components and takes values in the range $[-1,1]$. 
Note that, below, when referring to the ground state properties of the spinor system we present $\braket{\hat{P}(0)}\equiv \braket{\hat{P}}$ while for the time-evolution, see Section \ref{quench}, $\braket{\hat{P}(t)}$ is calculated. 
Thus regarding the above-described phases, i.e. F1, F2 and AF, it is anticipated that $\langle\hat{P}\rangle=-1$. 
This behavior is indeed confirmed both within the MF and in the beyond MF case as well as for different particle numbers as shown in Fig. \ref{fig:polar} (a), and essentially reflects the fact that $\braket{\hat{n}_0}=0$. 

Most importantly, in order to identify and subsequently quantify the corresponding phase transitions between the aforementioned magnetic phases, we further employ as an order parameter the expectation value of the spin-$z$ operator
\begin{equation} 
\begin{split}
\braket{\hat{F}_z(t)}=&\braket{\Psi(t)|\sum_{\alpha \beta}^{} \int dx \hat{\psi}_{\alpha}^\dagger(x) (f_z)_{\alpha \beta} \hat{\psi}_{\beta}(x) |\Psi(t)} \\& = \braket{\hat{n}_{1}(t)}-\braket{\hat{n}_{-1}(t)}. \label{spin_oparator}
\end{split}
\end{equation}
Note that $(f_z)_{\alpha\beta}=\alpha\delta_{\alpha\beta}$ denote the matrix elements of the spin-$z$ Pauli matrix whilst $\alpha,\beta \in \{-1,0,1\}$. 
The expectation value $\langle\hat{F}_z\rangle \in (-1,1)$ essentially measures the population imbalance between the $\alpha=1$ and $\alpha=-1$ components. 
Accordingly, $\langle\hat{F}_z\rangle=1$ and $\langle\hat{F}_z\rangle=-1$ indicate the occurrence of the F1 and the F2 phase respectively while if $-1<\langle\hat{F}_z\rangle <1$ then the AF phase is entered. 

\subsubsection{Transition from F2 to the P phase through the AF phase by adjusting the linear Zeeman term}

To be more precise, we focus on the existent second-order phase transition where the system transforms continuously from the F2 to the F1 phase via the AF phase as the linear Zeeman parameter $p/(c_1n)$ is increased for a fixed negative value of the quadratic Zeeman energy shift $q/(c_1n)$. 
Note that $n=\sum_\alpha^{}\langle \psi_\alpha^\dagger(x=0)\psi_\alpha(x=0)\rangle$ is the maximum value of the total  density at the trap center \footnote{For the systems examined in the main text the maximum value of the total spin density at the trap center takes the values $n\approx8.9$ for $N=500$, $n\approx1.9$ for $N=50$, $n\approx1$ for $N=20$ and $n\approx 0.4$ for $N=5$ in the dimensionaless units adopted herein.}. 
The behavior of $\langle \hat{F}_z\rangle$ for varying $p/(c_1n)$ and e.g. $q/(c_1n)=-0.5$ is presented in Fig. \ref{fig:spin}(a) both within the MF and the MB approach for different particle numbers. 
As it can be seen, the interval of $p/(c_1n)$ values where the AF state is accessed decreases in the presence of interparticle correlations. 
Also, $\langle \hat{F}_z\rangle$ acquires larger (smaller) values for $p/(c_1n)>0$ ($p/(c_1n)<0$) in the MB case. 
Indeed the transition point e.g. between AF and F1 is shifted towards $p/(c_1n)=0$ in the correlated case, compare in particular the MF and MB results for $N=5,20$ and $50$ particles in Fig. \ref{fig:spin}(a). 
Interestingly the deviation in the shape of $\langle \hat{F}_z \rangle$ between the MF and the MB approach as a function of $p/(c_1n)$ is more prominent in the few-body case e.g. $N=5$ in Fig. \ref{fig:spin} (a) and becomes smaller for a larger particle number e.g. $N=50$. 
The latter suggests that when approaching the thermodynamic limit i.e. $N\to \infty$ the difference of the MF to the MB result will be negligible at least for the considered ratio of spin-dependent over spin-independent interaction strengths i.e. $c_1/c_0=0.036$. 
Also, within the AF phase $\langle \hat{F}_z \rangle$ increases almost linearly for a larger $p/(c_1n)$ irrespectively of $N$. 
Furthermore, we can deduce that the transition threshold between the AF and F1 phases (or equivalently among the AF and the F2 phase) in terms of $\abs{p}/(c_1n)$ decreases for a smaller particle number, see $\langle \hat{F}_z \rangle$ for $N=5$, $N=20$ and $N=50$ bosons. 

Another interesting observation here is that the phase transition boundary which is known to occur at $p/(c_1n)=\pm1$ \cite{kawaguchi2012spinor,stamper2013spinor} for $N\to \infty$, it takes place at $p/(c_1n)<1$ due to the finite size effects emanating in our system, see for instance that for $N=5$ bosons in Fig. \ref{fig:spin} (a) the transition occurs at $p/(c_1n)=0.6$ in the MF limit. 
To verify the presence of finite size effects, we present $\langle \hat{F}_z \rangle$ for $N=500$ bosons within the MF approximation for varying $p/(c_1n)$ in the inset of Fig. \ref{fig:spin} (a). 
Evidently, in this case the transition indeed occurs at $p/(c_1n) \approx 1$, thus confirming that the transition point approaches $p/(c_1n)=1$ as $N\to \infty$. 
This, together with the independence of the results on $q/(c_1 n)$, suggests
that the phase diagram of the left panel of Fig.~\ref{fig:phase_dia}
retains its qualitative form, yet with the AF state suppressed as we go
to smaller $N$ and more so (as $N$ is lowered) in the MB case in comparison
with the MF one.

\subsection{Ferromagnetic ensembles} \label{ferro_int}

Turning to ferromagnetic interactions, i.e. $c_1<0$, there are three distinct second-order quantum phase transitions, see Fig. \ref{fig:phase_dia} (b) regarding $N \to \infty$. 
The involved phases correspond to the F1, F2, P and BA phases. 
Since the P phase consists of bosons being entirely in the $\alpha=0$ component, it holds that $\langle \hat{P}\rangle=1$ [Fig. \ref{fig:polar}(b)]. 
However, in the BA phase all the three spin-states are occupied and therefore $\langle \hat{P} \rangle \in(-1,1)$. 

\subsubsection{Transition from F2 to the P phase via the BA phase in terms of the quadratic Zeeman term} 

Figure \ref{fig:spin}(b) illustrates $\langle \hat{F}_z \rangle$ with respect to $q/(|c_1|n)$ for a specific $p/(|c_1|n)=-0.4$. 
The transition from the F2 [$\langle\hat{F}_z\rangle=-1$] to the P [$\langle\hat{F}_z\rangle=0$] phase via a monotonous increase of $\langle \hat{F}_z \rangle$ for larger $q/(|c_1|n)$ through the BA phase takes place. 
In more detail, the phase transition between the F2 and BA phases is expected to occur at $p=-q$ for $N\to \infty$~\cite{kawaguchi2012spinor,stamper2013spinor}. 
This fact is explicitly verified within our calculations both at and beyond the MF approximation as well as for different particle numbers [Fig. \ref{fig:spin}(b)]. 
Furthermore, the transition from the BA to the P phase exhibits a similar behavior to the one between the AF and F2 phases discussed in the anti-ferromagnetic ($c_1>0$) case [Fig. \ref{fig:spin}(a)]. 
Indeed, the transition point is negatively shifted to smaller values of $q$ in the MB case, an effect which is more pronounced in the few-body scenario. 
Additionally, $\langle \hat{F}_z \rangle$ within the BA phase exhibits larger values for fixed $p$, $q$ when correlations are present. 
This is more evident especially for smaller particle numbers, e.g. compare $\langle \hat{F}_z \rangle$ obtained in the MF and MB case for $N=5$ and $N=50$ in Fig. \ref{fig:spin}(b). 
The aforementioned overall phenomenology regarding the behavior of $\langle \hat{F}_z \rangle$ in the BA and P phases is also imprinted in the shape of $\langle \hat{P} \rangle$ due to the non-zero occupation of the $\alpha=0$ component in these phases, see Fig. \ref{fig:polar}(b). 

Concluding, we note that the transition from the BA to the P phase is expected to occur at $q/(|c_1|n)=2.07$ in the thermodynamic limit
(for this value of $p/(|c_1| n)$)~\cite{kawaguchi2012spinor,stamper2013spinor}, see also the behavior of this transition boundary in Fig. \ref{fig:phase_dia}(b). 
This is not observed in our system owing to the presence of finite size effects. 
To support this argument, we showcase in the inset of Fig. \ref{fig:spin}(b) $\langle \hat{F}_z \rangle$ for $N=500$ particles in MF limit i.e. when accounting for the hybridization between the spatial and spin degrees-of-freedom of the system.  
It becomes evident that in the $N=500$ case the transition takes place at $q/(|c_1|n) \approx 2.07$, thus confirming that the above-mentioned behavior is indeed a finite size effect. 
Moreover, within the MF approximation it is anticipated that the P phase is only accessed for $q/(|c_1|n)>2$ while in our case it is already realized for $q/(|c_1| n)=1.8$. 
To understand whether this behavior is a consequence of the finite particle number we have calculated $\langle \hat{P} \rangle$ in the MF limit for $N=500$ particles and $p/(|c_1|n)=0$ for varying $q/(|c_1|n)$, see the inset of Fig. \ref{fig:polar}(b). 
It can be easily seen that here the transition point to the P state is at $q/(|c_1|n)\approx 2$ which agrees with the theoretical prediction and consequently confirms the presence of finite size effects. 

\subsubsection{Transitions between the F1, BA, P and the F2 phases with varying linear Zeeman field} 

Next, we proceed by inspecting the properties of the second-order quantum phase transitions of a ferromagnetically interacting $c_1<0$ Bose gas taking place for fixed $q/(|c_1|n)=1.8$ and varying $p/(|c_1|n)$, see Fig. \ref{fig:spin}(c) and
Fig. \ref{fig:polar}(c). 
Since in this case several second-order transitions are in play in the thermodynamic limit \cite{kawaguchi2012spinor,stamper2013spinor} as also depicted in Fig. \ref{fig:phase_dia}(b), below we distinguish between the $p>0$ and the $p<0$ scenaria. 
For decreasing $p$ such that always $p>0$, the system transits from the F1 [$\langle\hat{F}_z\rangle=1$, $\langle \hat{P}\rangle=-1$] to the BA [$\langle \hat{F}_z \rangle\in(-1,1)$, $\langle \hat{P} \rangle \in(-1,1)$] phase and subsequently to the P [$\langle \hat{F}_z \rangle=0$, $\langle \hat{P}\rangle=1$] one. 
These emergent transitions among the above-described phases are indeed imprinted in the shape of $\langle\hat{F}_z\rangle$ and $\langle \hat{P}\rangle$ shown in Fig. \ref{fig:spin}(c) and Fig. \ref{fig:polar}(c) respectively. 
As it can be readily seen, $\langle\hat{F}_z\rangle$ [$\langle \hat{P}\rangle$] decreases [increases] for smaller positive values of $p$ independently of the MF or the MB case and the considered particle number. 
However, $\langle\hat{F}_z\rangle$ [$\langle \hat{P}\rangle$] is reduced [enhanced] in the MB compared to MF scenario while the transition point to the P phase gets positively shifted in the MB case, a result which is more prominent for decreasing atom number e.g. see $N=5$ and $N=20$ in Fig. \ref{fig:spin}(c) [Fig. \ref{fig:polar}(c)].

Entering the range of smaller $p$ values all the way to $p<0$, $\langle \hat{F}_z \rangle$ and $\langle \hat{P}\rangle$ decrease since the system moves first from the P to the BA phase and then from the BA to the F2 [$\langle \hat{F}_z \rangle=-1$, $\langle \hat{P}\rangle=-1$] phase, see Fig. \ref{fig:spin}(c) and Fig. \ref{fig:polar}(c). 
These transitions occur both in the MF and the MB approach and for different number of bosons. 
Evidently, the inclusion of correlations causes a significant deviation in both $\langle \hat{F}_z \rangle$ and $\langle \hat{P}\rangle$ within the BA phase when compared to the MF approximation. Indeed, these observables acquire smaller values and the transition from the P to the BA phase is negatively shifted closer to $p=0$ in the MF compared to the MB case. 
Therefore, the BA phase is effectively suppressed within the parametric plane, see e.g. Fig. \ref{fig:spin}(c), when correlations are taken into account. 
For instance, for $N=5$ bosons the P to the BA phase transition point takes place at $p/(|c_1|n)=\pm 1.44$ in the MB case and at $p/(|c_1|n)=\pm 0.96$ in the MF limit. 
This phenomenon is more pronounced for fewer particles, e.g. compare $\langle \hat{F}_z \rangle$ and $\langle \hat{P}\rangle$ for $N=5$ and $N=20$, and becomes vanishingly small for increasing $N$ see for instance the case of $N=50$. 
In addition, a smaller number of bosons also extends the $p$ interval in which the P phase exists, e.g. contrast $\langle \hat{F}_z \rangle$ when $N=5,20$ and $N=50$ in Fig. \ref{fig:spin}(c). 
This behavior can in turn be attributed to the finite size of the system. 
To explicitly visualize this fact we consider an adequately large particle number namely $N=500$ with fixed $q/(|c_1|n)=1.8$ and illustrate the behavior of $\langle \hat{F}_z \rangle$ [$\langle \hat{P}\rangle$] in the inset of Fig. \ref{fig:spin}(c) [Fig. \ref{fig:polar}(c)]. 
It is apparent that the shapes of $\langle \hat{F}_z \rangle$ and $\langle {\hat{P}} \rangle$ become much smoother within the P phase whose $p$ interval shrinks accordingly and thus approaches its expected thermodynamic behavior, see also Fig. \ref{fig:phase_dia}(b). 
We finally remark that the transition between the F1 and the BA [BA and F2] phases when $N \to \infty$ is expected \cite{kawaguchi2012spinor} to occur (as also shown in Fig. \ref{fig:phase_dia}(b) for $q/(|c_1|n)>2$) at $p=q$ [$p=-q$]. 
This result is indeed confirmed within our calculations and remains unaltered in both the MF and MB case as shown e.g. in Fig. \ref{fig:spin}(c). 

\subsubsection{Transitions among the F1, BA, and the F2 phases by tuning the linear Zeeman term} 

Subsequently, we turn our attention to the second-order quantum phase transition from the F2 to the F1 phase through the BA one exhibited in the thermodynamic limit upon increasing $p/(|c_1|n)$  for negative spin-dependent interactions $c_1<0$, see also Fig. \ref{fig:phase_dia}(b). 
To this end, we consider a fixed $q/(|c_1|n)=0.8$ and $c_1<0$. 
The population imbalance among the involved spin-states as quantified via $\langle \hat{F}_z \rangle$ and $\langle \hat{P}\rangle$ is presented in Fig. \ref{fig:spin}(d) and Fig. \ref{fig:polar}(d) respectively for varying $p/(|c_1|n)$ both at and beyond the MF level and for different numbers of bosons. 
In the thermodynamic limit \cite{kawaguchi2012spinor,stamper2013spinor} the phase transition between the ferromagnetic phase F1 [F2] and the BA one occurs at $p=q$ [$p=-q$]. 
More specifically, we observe that within the BA phase $\langle \hat{F}_z \rangle$ decreases monotonously for decreasing $p/(|c_1|n)$ whilst $\langle \hat{P}\rangle$ grows [reduces] for a smaller $p/(|c_1|n)$ such that $p>0$ [$p<0$]. 
In accordance with the above-discussed second-order transitions [Fig. \ref{fig:spin}(b), Fig. \ref{fig:spin}(c)], we can deduce that the underlying phase boundaries indeed occur at $p=\pm q$ and their positions are not significantly altered when correlations are taken into account [Fig. \ref{fig:spin}(d)]. 
Interestingly, in the correlated case and for $N=5$ bosons, the transition from the BA to the P phase and vice versa takes place at $p/(|c_1| n)=0.35$ and $p/(|c_1| n)=-0.35$ respectively [Fig. \ref{fig:spin} (d) and Fig. \ref{fig:polar} (d)]. 
As already discussed above, this effect can be understood due to the fact that the P phase is favorable for smaller values of $q/(|c_1| n)$ deep in the few-body regime e.g. $N=5$ and in the presence of correlations. 
For the remaining cases, there are minor changes in the value of either $\langle \hat{F}_z \rangle$ or $\langle \hat{P} \rangle$ within the BA phase obtained between the MF and the MB approach; it is relevant to compare in particular these observables in the BA region between the MB and the MF scenario for $N=20,50$ [Fig. \ref{fig:spin}(d), Fig. \ref{fig:polar}(d)]. 

Summarizing, we can deduce that the interval of the AF and the BA phases, with respect to the quadratic and the linear Zeeman energy terms, appearing in the phase diagram of a spin-1 Bose gas for positive ($c_1>0$) and negative ($c_1<0$) spin-dependent interactions respectively shrinks in the presence of intra- and intercomponent correlations. 
Moreover, this effect is more pronounced in the few-body case. 
As a consequence, one can infer that the phases involving a superposition of different spin-states are not favorable when  operating beyond the MF approximation, a phenomenon which becomes more evident for a decreasing particle number. 
Instead, single-component spin-states appear to be favored for lower particle
numbers and more so in the MB setting.
Finally, let us comment that the correlation patterns building upon the one- and two-body correlation functions of the system [see Eqs. (\ref{onebody_coherence}) and (\ref{twobody_coherence}) below] are similar to the ones appearing in the course of the evolution, see e.g. Fig. \ref{fig:cohcor}, and therefore we do not present them for the ground state. 
Moreover the inclusion of correlations leads to filamentary structures in the density of each 
component, an effect that is also evident during evolution see in particular the discussion below.

\section{Quench Dynamics}
\label{quench} 

Having analyzed in detail the correlation properties of the ground state phase diagram of the spin-1 Bose gas for different particle numbers, we then discuss some basic correlation aspects of its nonequilibrium dynamics. 
To this end, we investigate the dynamics induced by a sudden decrease (quench) of the harmonic trap frequency of the spin-1 Bose gas consisting of $N=50$ particles both within and beyond the MF approximation. 

More specifically, the system is initialized in its ground state configuration characterized by a fixed spin-dependent (spin-independent) interaction strength $c_1=-0.0047 \sqrt{\hbar^3\omega_\perp/M}$ ($c_0=1\sqrt{\hbar^3\omega_\perp/M}$) for ferromagnetic interactions and $c_1=0.018 \sqrt{\hbar^3\omega_\perp/M}$  ($c_0=0.5\sqrt{\hbar^3\omega_\perp/M}$) for anti-ferromagnetic couplings (unless it is stated otherwise) but a variety of Zeeman parameters $p$, $q$. 
The latter allows us to enter the different ground state phases discussed in Section \ref{phase_space}. 
These include for instance the F1, F2, AF, BA and P phases (see for more details below). 
To trigger the dynamics we perform a quench of the trapping frequency from $\omega=0.1$ to $\omega=0.07$ and monitor the emergent dynamical response of the individual components utilizing their single-particle density and associated correlation functions. 
Naturally, this quench protocol excites a collective breathing motion \cite{breathing_1,pyzh2018spectral} of the individual components.

\begin{figure}[ht]
        \centering
            \includegraphics[width=0.47\textwidth,height=\textheight,keepaspectratio]{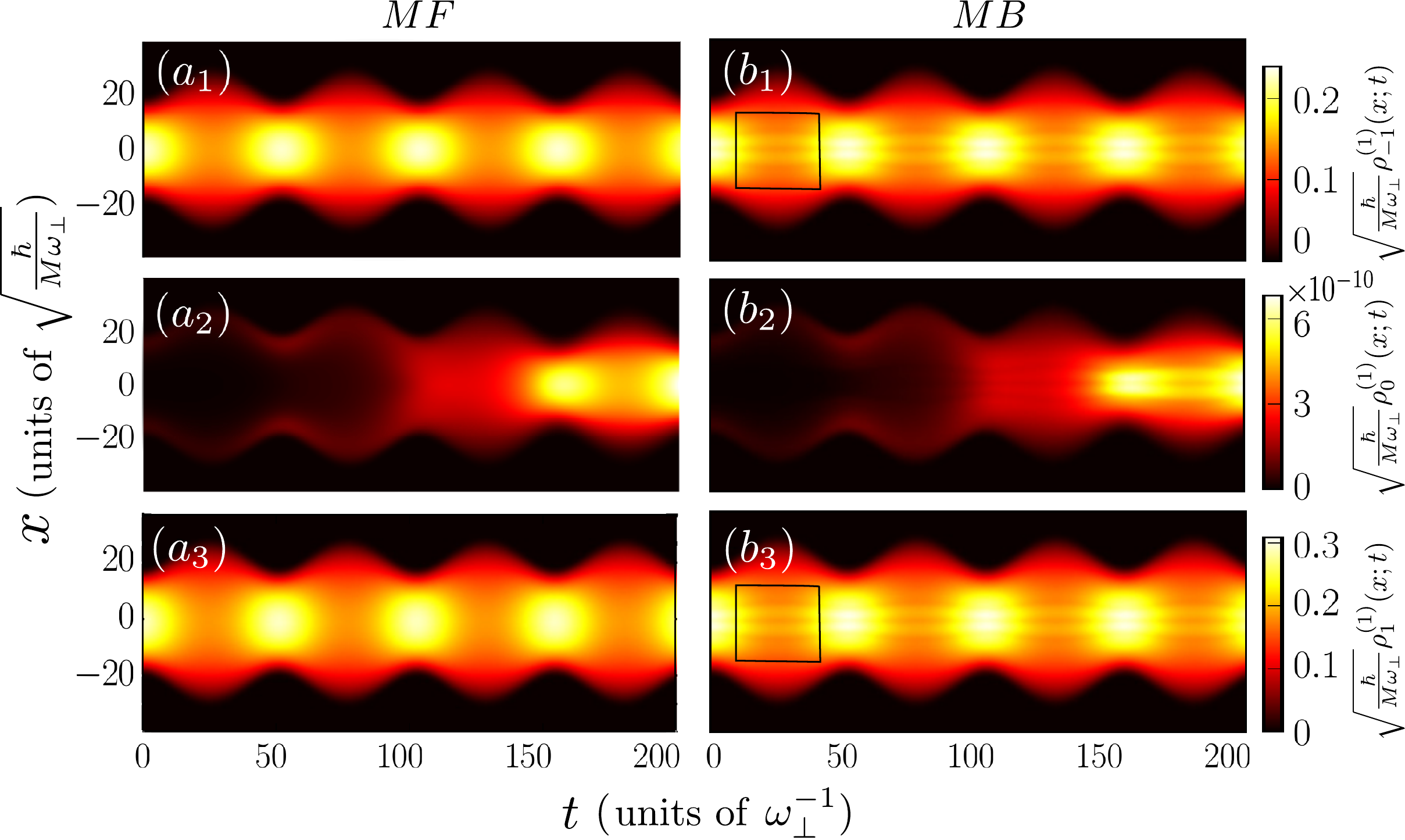}
        \caption{Spatiotemporal evolution of the spin-resolved one-body densities $\rho^{(1)}_{\alpha}(x;t)$ of the spin-1 Bose gas in an AF phase characterized by Zeeman energy parameters $p/(c_1n)=0.11$ and $q/(c_1n)=-0.5$ within the (a) MF and (b) MB approach. 
        Panels correspond to the ($a_1$), ($b_1$) $\alpha=-1$, ($a_2$), ($b_2$) $\alpha=0$ and ($a_3$), ($b_3$) $\alpha=1$ components (see legends). 
        The harmonically trapped spin-1 Bose gas consisting of $N=50$ particles with spin-independent interaction strength $c_0=0.5\sqrt{\hbar^3\omega_\perp/M}$ and spin-dependent one $c_1=0.018 \sqrt{\hbar^3\omega_\perp/M}$ is initially prepared in its ground state. 
        To induce the dynamics a quench of the trapping frequency from $\omega=0.1$ to $\omega=0.07$ is applied at $t=0$. 
        Note that the color scaling is different for each component; in particular,
        note the fundamentally different scale of the $\alpha=0$ component.} 
        \label{fig:quench}
\end{figure}

\subsection{One-body density evolution}\label{density_evolution} 

\begin{figure*}[ht]{}
    \includegraphics[width=0.8\textwidth,height=\textheight,keepaspectratio]{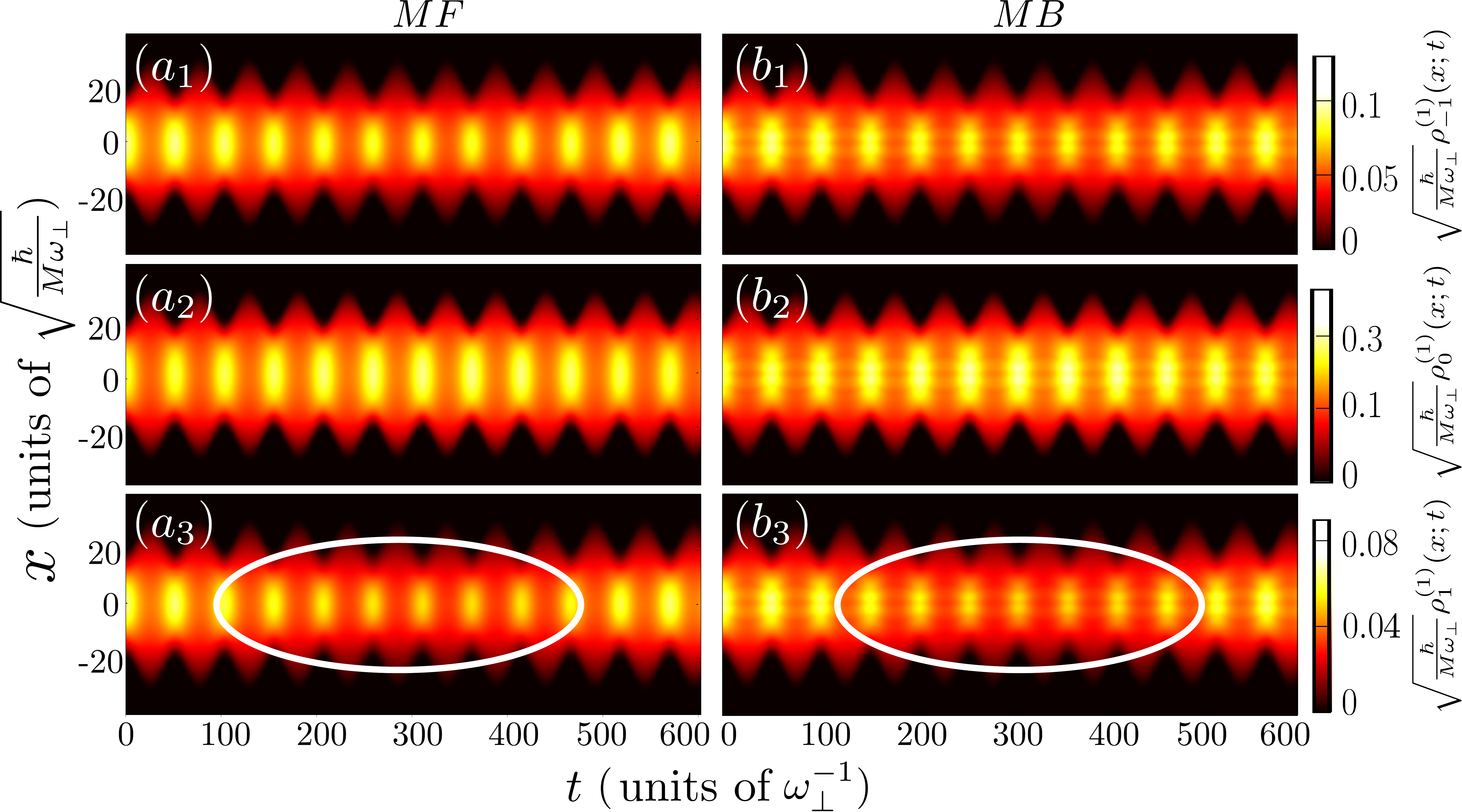}
    \caption{Time-evolution of the spin-resolved one-body densities $\rho^{(1)}_{\alpha}(x;t)$ within the (a) MF and (b) MB approach of the spin-1 Bose gas in a BA phase with Zeeman parameters $p/(|c_1|n)=0.04$ and $q/(|c_1|n)=0.44$.
    The individual panels illustrate the ($a_1$), ($b_1$) $\alpha=-1$, ($a_2$), ($b_2$) $\alpha=0$ and ($a_3$), ($b_3$) $\alpha=1$ spin-states. 
    The spin-1 Bose gas with $N=50$ atoms is harmonically trapped and it is initialized in its ground state configuration. 
    The spin-independent and spin-dependent interaction strengths are $c_0=1 \sqrt{\hbar^3\omega_\perp/M}$ and $c_1=-0.0047 \sqrt{\hbar^3\omega_\perp/M}$ respectively.   
    To trigger the dynamics we follow a quench of the trapping frequency from $\omega=0.1$ to $\omega=0.07$ at $t=0$.}
    \label{fig:quench_5}
\end{figure*}

To visualize the emergent nonequilibrium dynamics of the spin-1 Bose gas we first track the time-evolution of the single-particle density $\rho_{\alpha}^{(1)}(x;t)=\langle \Psi(t)|\hat{\psi}^{\dagger}_{\alpha}(x)\hat{\psi}_{\alpha}(x)|\Psi(t)\rangle$ of each component \cite{lode2016multiconfigurational}. 
Prototypical examples of the induced dynamics are shown in Figure \ref{fig:quench} and Figure \ref{fig:quench_5} demonstrating $\rho_\alpha^{(1)}(x;t)$ with $\alpha=-1,0,1$ following a quench of the harmonic trap frequency within the AF  and the BA phase respectively, which emanate in our system for ferromagnetic $c_1<0$ and anti-ferromagnetic $c_1>0$ spin-spin interactions. 
As it can be readily seen in both cases this quench protocol results in an induced breathing motion \cite{breathing_1,pyzh2018spectral}, manifested as a contraction and expansion dynamics, of the atomic cloud of the individual components around the trap center. 
Interestingly, the frequency of the breathing oscillation remains unaltered, namely $\omega_{br}^{\alpha}=0.12$, regarding the bosons residing in different spin-states [e.g. compare Figs. \ref{fig:quench} ($b_1$), ($b_3$)], between the MF and the MB case [see for instance Figs. \ref{fig:quench} ($a_1$),($b_1$)] as well as irrespectively of the initial phase of the spinor gas [see e.g. Figs. \ref{fig:quench}($b_1$), ($b_3$) and Figs. \ref{fig:quench_5}($b_1$), ($b_3$)]. 
Note here that the deviation of the observed breathing frequency, $\omega_{br}^{\alpha}=0.12$, from its theoretically anticipated value i.e. $\omega_{br;th}^{\alpha}=0.14$ is attributed to the finite size of the system \cite{breathing_1,pyzh2018spectral}.   
A notable effect of the presence of correlations on the single-particle level is the formation of filament-like structures building upon the one-body densities of the individual spin-components as depicted in Fig. \ref{fig:quench} and Fig. \ref{fig:quench_5}. 
This filamentary configuration refers to the multihump patterns appearing in the density of the Bose gas and being more prominent during its expansion dynamics, see e.g. the box in Fig. \ref{fig:quench}. 
Such a type of filamentation process is the imprint of correlations in $\rho_\alpha^{(1)}(x;t)$ and has already been observed in correlated binary bosonic \cite{mistakidis2018correlation,mistakidis2020many} and fermionic \cite{erdmann2019phase,kwasniok2020correlated} mixtures. 
\begin{figure}[ht]{}
    \centering
    \includegraphics[scale=0.18]{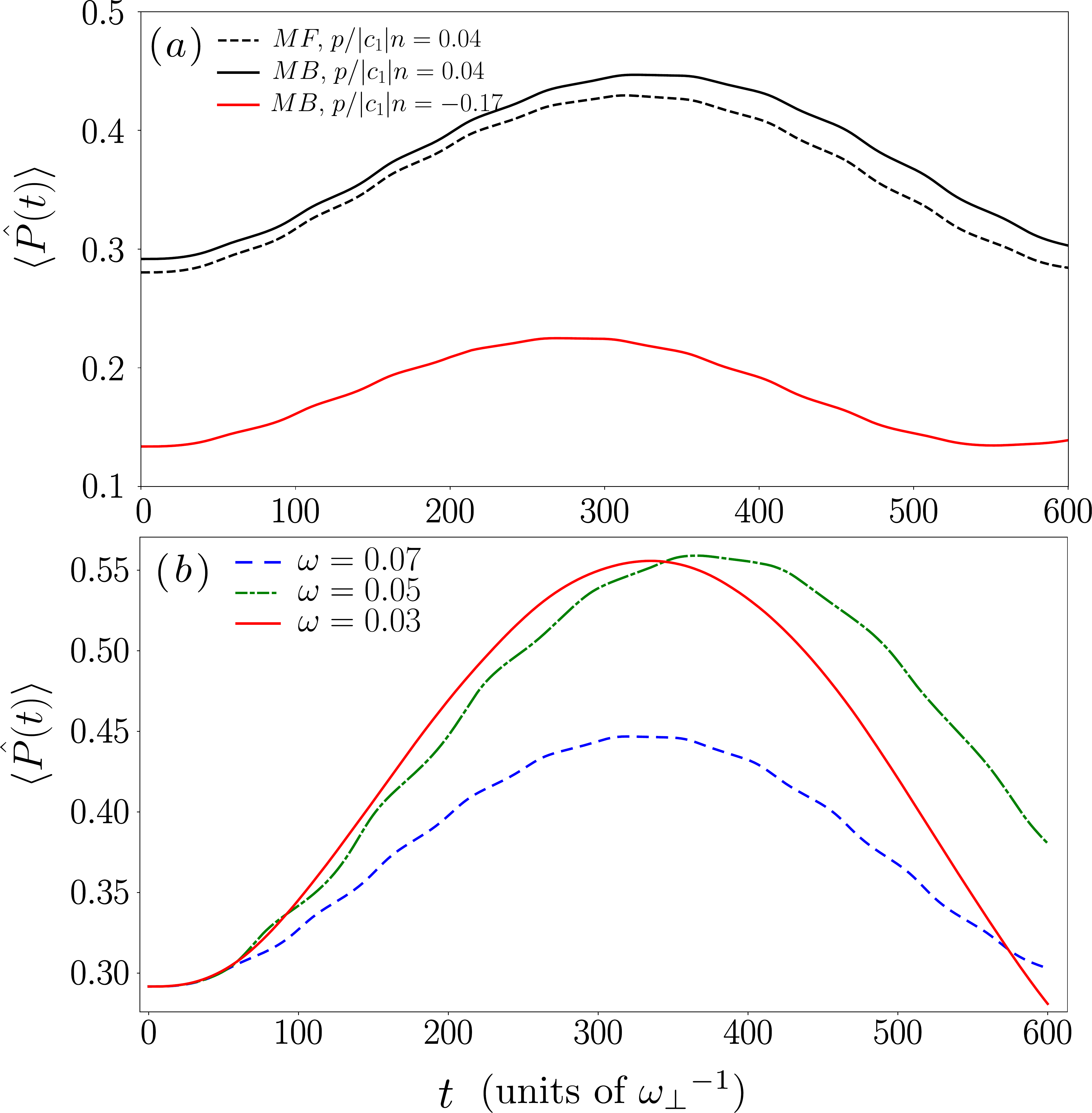}
    \caption{(a) Time-evolution of the population imbalance, $\langle \hat{P} (t) \rangle$, between the $\alpha=0$ and the $\alpha=\pm 1$ components, after a quench of the harmonic trap frequency from $\omega=0.1$ to $\omega=0.07$. 
    Solid lines denote $\langle \hat{P} (t) \rangle$ within the MB approach while the dashed line 
    represents the MF case. 
    (b) The same as in (a) with $p/(|c_1|n)=0.04$ but for quenching the trap frequency from $\omega=0.1$ to different values of $\omega$ (see legend). 
    The system is initialized in its ground state residing in the BA phase with $q/(|c_1|n)=0.44$ and different $p/(|c_1|n)$ values (see legend).
    Other system parameters are the same as in Fig. \ref{fig:quench_5}.}
    \label{fig:polar_quench}
\end{figure}

\subsection{Spin-mixing processes} 

Another crucial observation here is that quenching an initially AF state does not lead to a significant spin-flip dynamics among the participating components. 
Indeed, the probability of particles lying in the distinct spin-states remains almost constant in the course of the evolution, while the total spin of the system is conserved at each time instant. 
Similar observations in terms of the negligible spin-flip dynamics and the independence of the breathing frequency can also be drawn for an initially ferromagnetic either F1 or F2 phase as well as the P one (not shown for brevity). 
In sharp contrast to the above, when a BA phase is subjected to such a quench a low frequency spin-flip dynamics between the components occurs \cite{spin_fluctuations}, see Fig. \ref{fig:quench_5}. 
More specifically, at the initial stages of the dynamics the particles residing in the $\alpha=1$ and the $\alpha=-1$ components get coherently transferred towards the $\alpha=0$ spin-state while the reverse scenario is subsequently realized, see for instance the region indicated by an ellipse in Fig. \ref{fig:quench_5} corresponding to the dynamics of the $\alpha=1$ component. 

To quantify this population transfer among the $\alpha=0$ and $\alpha =\pm 1$ spin-states we consequently monitor the time-evolution of the expectation value of the underlying polar operator i.e. $\langle \hat{P}(t) \rangle$ after a quench of the harmonic trap frequency to $\omega=0.07$. 
The dynamics of this observable within the BA phase is presented in Fig. \ref{fig:polar_quench}(a) for Zeeman energy parameters $q/(|c_1|n)=0.44$ and $p/(|c_1|n)=0.04$ within the MF and the MB evolution as well as for $p/(|c_1|n)=-0.17$ only in the MB approach. 
Focusing on the case of $p/(|c_1|n)=0.04$ we observe that $\langle \hat{P}(t) \rangle$ increases until $t\approx300$, implying a transfer of bosons from the $\alpha = \pm 1$ to the $\alpha=0$ spin-state, and later on exhibits a decreasing behavior testifying a reverse migration tendency of the atoms namely $\alpha=0 \to \alpha= \pm 1$ in both approaches. 
The amplitude of this transfer process is somewhat larger in the presence of correlations, see also the corresponding behavior of $\langle \hat{P} (t=0) \rangle$ in the ground state at and beyond the MF level [Fig. \ref{fig:polar} (b)]. 
Also, referring to the MB evolution the migration of bosons between the aforementioned spin-states occurs faster upon decreasing the strength of the linear Zeeman energy term and also a smaller amount of particles is transferred to the $\alpha=0$ component, see Fig. \ref{fig:polar_quench}(a). 
Moreover, the amplitude of this transmission process increases for a larger quench amplitude and fixed $p$, $q$ parameters at least for $t<350$. 
For instance, compare $\langle \hat{P}(t) \rangle$ for the different post-quench harmonic trap frequencies $\omega=0.07$ and $\omega=0.03$ illustrated in Fig. \ref{fig:polar_quench}(b) when $p/(|c_1|n)=0.04$, $q/(|c_1|n)=0.44$. 
Notice, however, that for long evolution times here $t>350$ there is no clear categorization of the amplitude of $\langle \hat{P}(t) \rangle$ with respect to the post-quench $\omega$. 
Indeed, $\langle \hat{P}(t) \rangle$ is larger (smaller) for $\omega=0.03$ than for $\omega=0.05$ when $t<350$ ($t>350$). 
We should also stress at this point that in all of the above-mentioned cases $\langle \hat{F}_z \rangle$ remains constant throughout the time-evolution meaning that the overall population imbalance between the $\alpha=1$ and $\alpha=-1$ spin-states is not affected by the quench. 
Summarizing, we can infer the control of the intercomponent transfer process within the BA phase via tuning either the amplitude of the linear Zeeman parameter [Fig. \ref{fig:polar_quench}(a)] or the quench amplitude [Fig. \ref{fig:polar_quench}(b)]. 
\begin{figure}[ht]
        \includegraphics[width=0.47\textwidth,height=\textheight,keepaspectratio]{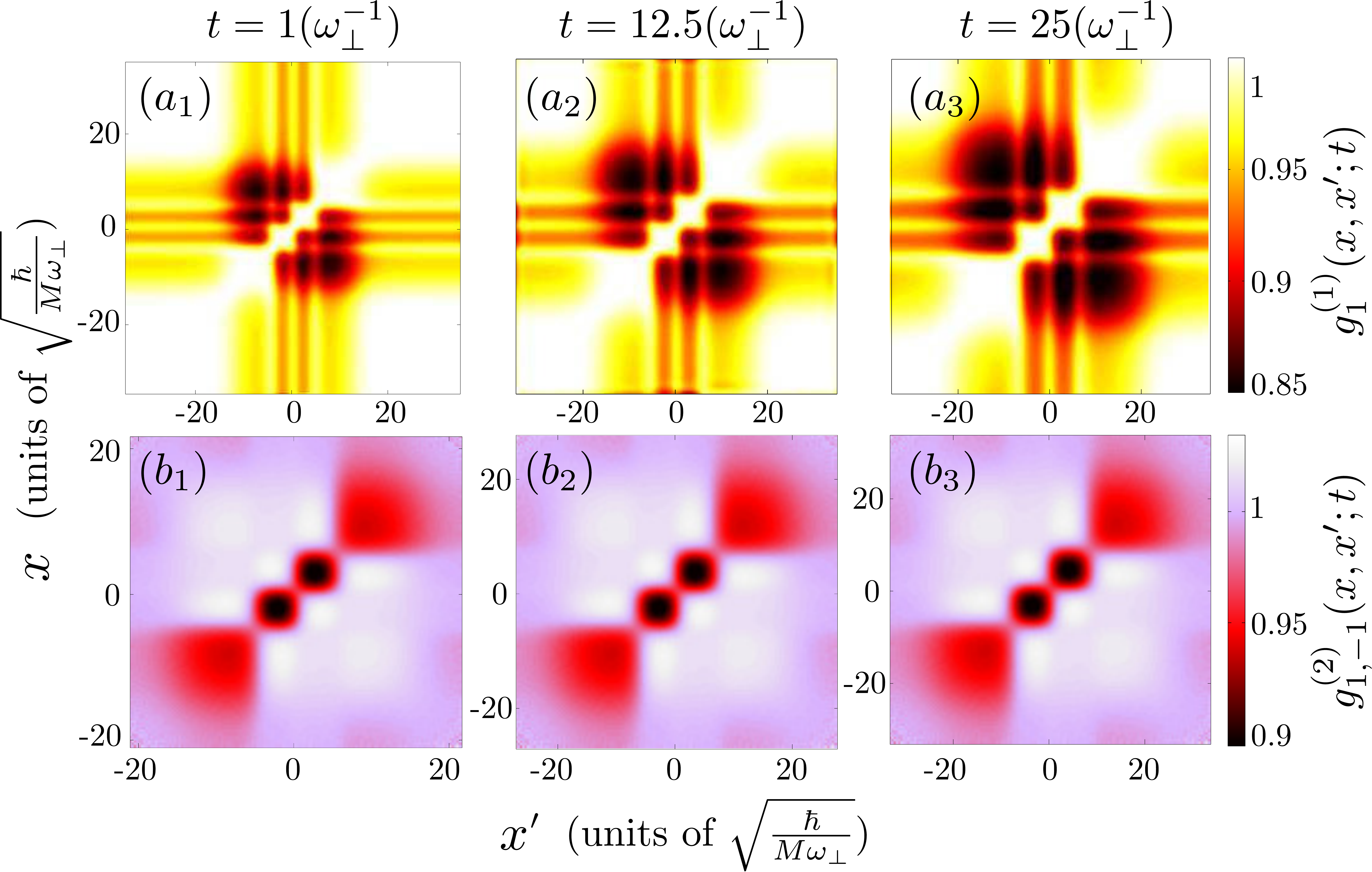}
          \caption{($a_1$), ($a_2$), ($a_3$) Snapshots of the one-body coherence function $g^{(1)}_{1}(x,x';t)$ of the $\alpha=+1$ component of the spin-1 Bose gas residing in the AF phase as described in Fig. \ref{fig:quench}. 
          ($b_1$), ($b_2$), ($b_3$) The corresponding two-body correlation function $g^{(2)}_{1,-1}(x_1,x_2;t)$ between the $\alpha=+1$ and $\alpha=-1$ components at different time-instants of the evolution. 
          The harmonically trapped spinor gas with $N=50$ bosons is prepared in its ground state characterized by $c_0=0.5\sqrt{\hbar^3\omega_\perp/M}$ and $c_1=0.018 \sqrt{\hbar^3\omega_\perp/M}$. 
          The dynamics is induced by applying a quench of the trapping frequency from $\omega=0.1$ to $\omega=0.07$ at $t=0$. 
          All other parameters are the same as in Fig. \ref{fig:quench}.}
     \label{fig:cohcor}
\end{figure}

\subsection{Coherence properties}\label{coherence}

In an attempt to further expose the role of correlations during the breathing dynamics of the spin-1 Bose gas we next invoke the $\alpha=\pm 1, 0$ component spatially resolved first-order coherence function \cite{coh,naraschewski1999spatial}
\begin{equation}
    g^{(1)}_{\alpha}(x,x';t)= \frac{ \rho^{(1)}_{\alpha}(x,x';t)}{\sqrt{\rho^{(1)}_{\alpha}(x;t) \rho^{(1)}_{\alpha}(x';t)}}.\label{onebody_coherence}
\end{equation}
In this expression, $\rho_\alpha^{(1)}(x,x';t)=\langle \Psi(t)|\hat{\psi}^{\dagger}_\alpha(x)\hat{\psi}_\alpha(x')|\Psi(t)\rangle$ denotes the $\alpha$-component one-body reduced density matrix whose diagonal is the previously discussed one-body density, namely $\rho_\alpha^{(1)}(x,x'=x;t)=\rho_\alpha^{(1)}(x;t)$. 
Evidently, $g^{(1)}_{\alpha}(x,x';t)$ can be used to infer the proximity of the MB state to a MF (product) one for a fixed set of coordinates $x$, $x'$.  
\begin{figure}[ht]
        \includegraphics[width=0.47\textwidth,height=\textheight,keepaspectratio]{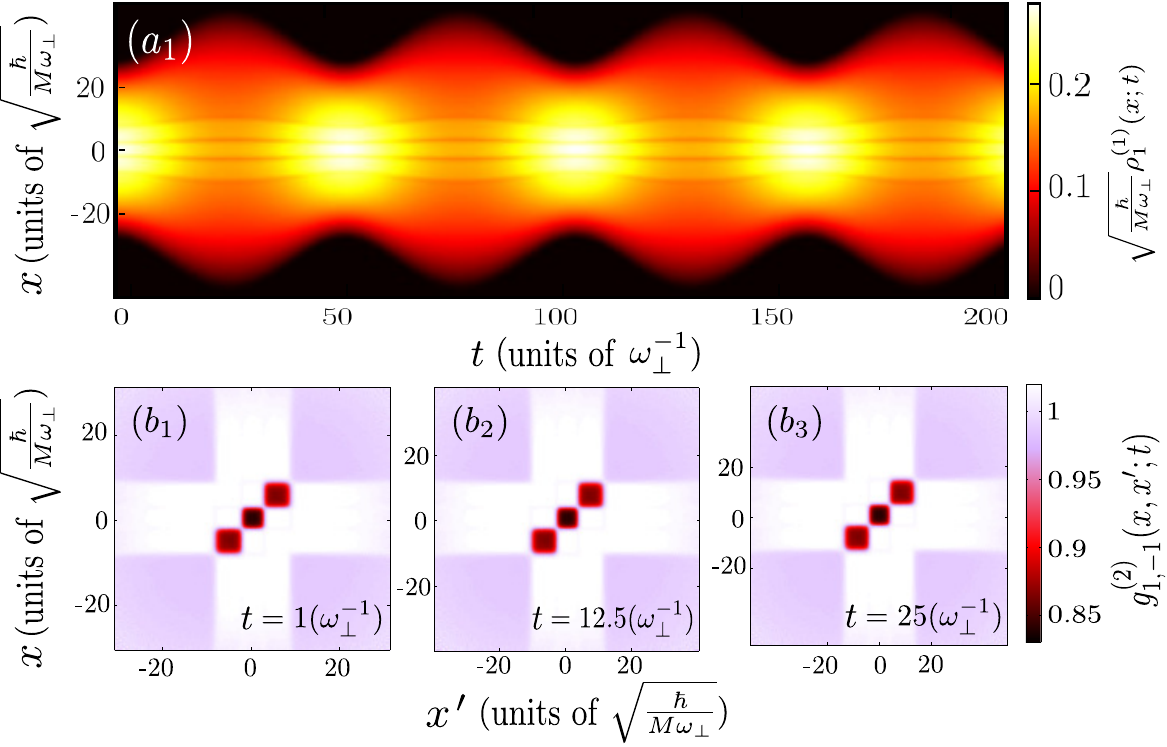}
          \caption{($a_1$) Spatiotemporal evolution of the one-body density $\rho^{(1)}_{1}(x;t)$ of the $\alpha=1$ component of the spin-1 Bose gas, being in the AF phase, within the MB approach. 
          ($b_1$), ($b_2$), ($b_3$) The two-body intercomponent correlation function $g^{(2)}_{1,-1}(x,x';t)$ at different time instants of the dynamics (see legend).
          The gas consists of $N=50$ particles with spin-independent interaction strength $c_0=5\sqrt{\hbar^3\omega_\perp/M}$ and spin-dependent one $c_1=0.018\sqrt{\hbar^3\omega_\perp/M}$ while confined in a harmonic trap. 
          It is initialized in its ground state where the Zeeman parameters are $p/(c_1n)=0.04$ and $q/(c_1n)=-0.44$.  
          The dynamics is triggered via a quench of the trapping frequency from $\omega=0.1$ to $\omega=0.07$.}
     \label{fig:c0_5}
\end{figure}
Additionally, $|g^{(1)}_{\alpha}(x,x';t)| \in [0,1]$ with $|g^{(1)}_{\alpha}(x,x';t)|=0$ ($|g^{(1)}_{\alpha}(x,x';t)|=1$) referring to a fully incoherent (coherent) state and thus indicating a maximal (zero) departure from the MF state. 
Accordingly, the absence of correlations is realized when $|g^{(1)}_{\alpha}(x,x';t)|=1$ for every $x$, $x'$ while the case of partial incoherence, i.e. $|g^{(1)}_{\alpha}(x,x';t)| < 1$ between two distinct spatial regions signifies the presence of correlations in the $\alpha$-component \cite{katsimiga2017dark}. 

Below, we analyze the coherence properties in the course of the breathing motion of the spin-1 Bose gas with $N=50$ atoms which is prepared in its ground state configuration where $p/(c_1n)=0.11$, $q/(c_1n)=-0.5$, $c_0=0.5\sqrt{\hbar^3\omega_\perp/M}$ and $c_1=0.018 \sqrt{\hbar^3\omega_\perp/M}$. 
Recall that for this choice of parameters the system resides in an AF phase. 
We also remark that in this case we have observed that the structures building upon the one-body coherence function are the same independently of the $\alpha= \pm 1$ component while the $\alpha=0$ spin-state remains unoccupied throughout the evolution, see also Figs. \ref{fig:quench}($b_1$)-($b_3$). 
For this reason we explicitly focus on the dynamical response of the $\alpha=+1$ component. 

Accordingly, $|g^{(1)}_{1}(x,x';t)|$ for the AF phase is illustrated in
Figs.~\ref{fig:cohcor}($a_1$)-\ref{fig:cohcor} ($a_3$) at specific time instants of the evolution referring to the contraction and expansion of the $\alpha$-component bosonic cloud.   
Overall, weak losses of coherence during the dynamics can be immediately inferred since the off-diagonal elements of $|g^{(1)}_{1}(x,x'\neq x;t)|$ exhibit values in the range $[ 0.85,1 ]$. Moreover, the patterns appearing in the one-body coherence remain robust during the evolution besides the contraction [Fig. \ref{fig:cohcor}($a_1$)] and expansion [Figs. \ref{fig:cohcor}($a_2$)-($a_3$)] of the off-diagonal which essentially reflects the breathing motion of the atomic cloud [Fig. \ref{fig:quench}($b_1$)]. In particular, closely inspecting $|g^{(1)}_{1}(x,x';t)|$ it becomes evident that the filamentary patterns imprinted in the $\alpha=+1$ spin-state one-body density are fully coherent among themselves. 
This is manifested by the bright blocks along the diagonal of $|g^{(1)}_{1}(x,x';t)|$, e.g. $g^{(1)}_{1}(x=-2,x'=-2;t=1) \approx 1$ in Figs. \ref{fig:cohcor}($a_1$)-($a_3$). 
Most importantly, nearest neighbor as well as long-distant filaments exhibit a prominent loss of coherence as it can be directly seen by inspecting the off-diagonal of $|g^{(1)}_{1}(x,x'\neq x;t)|$. 
For instance $g^{(1)}_{1}(x=-2,x'=3;t=1) \approx 0.88$ in Fig. \ref{fig:cohcor} ($a_1$) and $g^{(1)}_{1}(x=10,x'=-10;t=12.5) \approx 0.85$. 
It is also worth mentioning here that the coherence patterns are similar to the above-described also when starting from a BA phase and therefore we refrain on discussing also this case.

\subsection{Two-body correlation dynamics}\label{two_body_cor} 

Subsequently in order to unveil the interplay of two-body correlations in the dynamics of the spinor Bose gas we resort to the normalized two-body correlation function \cite{naraschewski1999spatial,mistakidis2020many,mistakidis2018correlation} given by
\begin{equation}
    g^{(2)}_{\alpha \alpha'}(x,x';t)= \frac{ \rho^{(2)}_{\alpha \alpha'}(x,x';t)}{\sqrt{\rho^{(1)}_{\alpha}(x;t) \rho^{(1)}_{\alpha'}(x';t)}}.\label{twobody_coherence}
\end{equation}
Here, $\rho^{(2)}_{\alpha \alpha'}(x,x';t)=\langle\Psi(t)| \hat{\psi}_{\alpha} ^\dagger(x) \hat{\psi}_{\alpha'} ^\dagger (x')\hat{\psi}_{\alpha'}(x') \hat{\psi}_{\alpha} (x) |\Psi(t) \rangle$ refers to the diagonal two-body reduced density matrix which gives the probability of two bosons of the $\alpha$ and $\alpha'$ components to be simultaneously at positions $x$ and $x'$ respectively. 
A perfectly condensed MB state leads to $|g^{(2)}_{\alpha \alpha'}(x,x';t)|=1$ and it is termed two-body uncorrelated while if $|g^{(2)}_{\alpha \alpha'}(x,x';t)|<1$ [$|g^{(2)}_{\alpha \alpha'}(x,x';t)|>1$] it is said to be two-body correlated [anti-correlated], see also Refs. \cite{mistakidis2018correlation,lode2019multiconfigurational} for more details.  
\begin{figure}[ht]
\includegraphics[width=0.47\textwidth,height=\textheight,keepaspectratio]{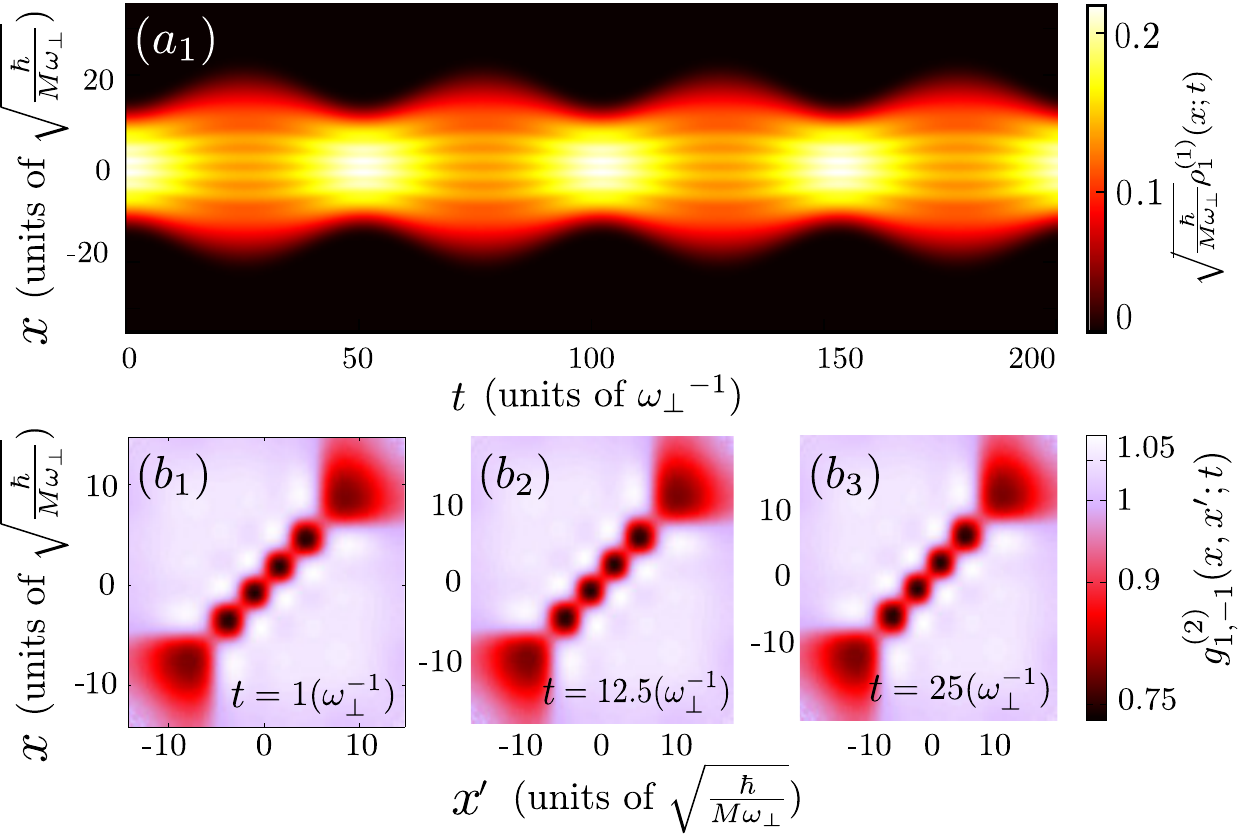}
\caption{ ($a_1$) Dynamics in the MB approach of the one-body density $\rho^{(1)}_{1}(x;t)$ of the $\alpha=1$ component of the spin-1 Bose gas in the AF phase.  
($b_1$), ($b_2$), ($b_3$) Profiles of the two-body intercomponent correlation function $g^{(2)}_{1,-1}(x,x';t)$ at specific time-instants of the time-evolution (see legend).
The harmonically trapped spin-1 Bose gas contains $N=20$ and characterized by spin-independent and spin-dependent interaction strengths $c_0=0.5\sqrt{\hbar^3\omega_\perp/M}$ and $c_1=0.018\sqrt{\hbar^3\omega_\perp/M}$ respectively.
It is prepared in the ground state with Zeeman terms $p/(c_1n)=0.04$ and $q/(c_1n)=-0.44$.  
To induce the nonequilibrium dynamics we apply a quench of the trapping frequency from $\omega=0.1$ to $\omega=0.07$.}
\label{fig:2050}
\end{figure}

Figures \ref{fig:cohcor}($b_1$)-($b_3$) present snapshots of the two-body intercomponent correlation function namely $g^{(2)}_{1,-1}(x,x';t)$ following a quench of the trapping frequency of the ground state prepared in the AF phase presented in Fig. \ref{fig:quench} and also discussed in section \ref{coherence}. 
Interestingly, for the dynamical response of our system $g^{(2)}_{\alpha \alpha'}(x,x';t)$ turns out to be almost independent of the involved components $\alpha=1$ and $\alpha'=-1$ while preserving its structure during the evolution. 
Recall that $\alpha=0$ is not populated. 
For the intracomponent two-body correlation function, of course, an expansion and contraction of the arising patterns, as an imprint of the overall $\alpha$-component breathing motion, takes place (not shown here for brevity) similarly to the one-body coherence function. 
Turning to $g^{(2)}_{1,-1}(x,x';t)$ we find a correlation hole to be present along its diagonal, see e.g. $g^{(2)}_{1,-1}(x=-2,x'=-2;t=1) \approx 0.9$ in Fig. \ref{fig:cohcor} ($b_1$), excluding the possibility an $\alpha=1$ and an $\alpha=-1$ boson  to reside in the same positioned filament \cite{mistakidis2018correlation}. 
This behavior is robust for all evolution times, i.e. two particles of different components lying in the same filament are anti-correlated, see the dark blocks near the diagonal in Figs. \ref{fig:cohcor} ($b_1$)-($b_3$). 
On the other hand, the off-diagonal elements show a weakly correlated character, thus allowing for an $\alpha=1$ and an $\alpha=-1$ particle to be at different filaments \cite{mistakidis2018correlation,erdmann2019phase,kwasniok2020correlated}, see in particular the bright white spots near the off-diagonal e.g. $g^{(2)}_{1,-1}(x=-2,x'=3;t=1) \approx 1.03$ in Fig. \ref{fig:cohcor}($b_1$). 

\subsection{Impact of the spin-independent interactions and the atom number on the dynamics} 

Let us finally examine the dependence of the above-discussed dynamics on the value of the spin-independent interaction strength and the number of atoms of the spin-1 Bose gas. 
For simplicity, we shall discuss only the case where the system in prepared in the ground state of an AF phase characterized by Zeeman parameters as shown in Fig. \ref{fig:quench}. 
Other phases such as F1, F1, P and BA yield similar observations and are not analyzed. 

As it can be readily seen in Fig. \ref{fig:c0_5} an increasing spin-independent interaction strength, e.g. $c_0=5\sqrt{\hbar^3\omega_\perp/M}$, leads to the appearance of relatively more prominent filamentary structures appearing in the corresponding one-body density of the $\alpha=+1$ component, compare Fig. \ref{fig:c0_5}($a_1$) and Fig. \ref{fig:quench}($b_1$). 
Furthermore, the number of filaments formed becomes also larger for a stronger $c_0$, e.g. from two when $c_0=0.5\sqrt{\hbar^3\omega_\perp/M}$ [Fig. \ref{fig:c0_5} ($b_3$)] to three for $c_0=5\sqrt{\hbar^3\omega_\perp/M}$ [Fig. \ref{fig:c0_5} ($a_1$)]. 
Note here that for brevity, we only show the density of the $\alpha=1$ spin-state, since the phenomenology of the $\alpha=-1$ component is similar while the $\alpha=0$ one is not populated due to the fact that the ground state lies in the AF phase, see also Fig. ~\ref{fig:quench}. 
On the two-body correlation level we showcase $g^{(2)}_{1,-1}(x,x';t)$ for $c_0=5\sqrt{\hbar^3\omega_\perp/M}$ at specific time-instants of the evolution in Figs. \ref{fig:c0_5}($b_1$)-($b_3$). 
A three block anti-correlated pattern appears along the diagonal of $g^{(2)}_{1,-1}(x,x';t)$, e.g. $g^{(2)}_{1,-1}(x=5,x'=5;t=1) \approx 0.85$ in Fig. \ref{fig:c0_5}($b_1$), with each block corresponding to a filament developing in the density of each component [Fig. \ref{fig:c0_5}($a_1$)]. 
We remark that this behavior is in contrast to the two block structure arising in $g^{(2)}_{1,-1}(x,x';t)$ for $c_0=1\sqrt{\hbar^3\omega_\perp/M}$
[Figs. \ref{fig:cohcor}($b_1$)-($b_3$)]. 
Additionally here the nearest neighboring, e.g. $g^{(2)}_{1,-1}(x=-5,x'=0.7;t=1)\approx1.1$, as well as the next-to-nearest neighboring filaments, e.g. $g^{(2)}_{1,-1}(x=-5,x'=5;t=1)\approx1.05$, are slightly correlated to each other, see also the off-diagonal bright region in Figs. \ref{fig:cohcor} ($b_1$)-($b_3$). 
Otherwise, the two-body correlation patterns possess the same characteristics as in the case of $c_0=0.5\sqrt{\hbar^3\omega_\perp/M}$. 

Next we consider the effect of a smaller particle number, $N=20$, while keeping fixed all other system parameters. 
Figure \ref{fig:2050}($a_1$) presents $\rho^{(1)}_{1}(x;t)$ where an increased number of filaments formed namely four occurs when compared to two for $N=50$ bosons [Fig. \ref{fig:quench}($b_1$)]. 
Also, $\rho^{(1)}_{1}(x;t)$ undergoes a breathing motion with the same frequency as in the $N=50$ scenario, a dynamical response that holds equally for the $\alpha=-1$ component whilst the $\alpha=0$ remains unoccupied. 
Recall here that the filamentary patterns constitute an imprint of the presence of correlations into the system and as such it is expected that their traces will be more pronounced for a decreasing particle number since in this latter case correlations are enhanced compared to larger particle systems \cite{mistakidis2018correlation}. 
Accordingly, the above-described behavior is also captured by the two-body correlation function [Figs. \ref{fig:2050}($b_1$)-($b_3$)]. 
Indeed, the occurrence of four prominent anti-correlated blocks along the diagonal of $g^{(2)}_{1,-1}(x,x'=x;t)$ is evident with each block corresponding to a particular filament. 
Note that this is again in contrast to the two anti-correlated block structure of $g^{(2)}_{1,-1}(x,x'=x;t)$ for $N=50$ particles [Figs. \ref{fig:cohcor}($b_1$)-($b_3$)]. 
Also, we can identify the presence of correlations between neighboring filaments, see the bright off-diagonal region of $g^{(2)}_{1,-1}(x,x';t)$ in Figs. \ref{fig:2050} ($b_1$)-($b_3$), e.g. $g^{(2)}_{1,-1}(x=-1.7,x'=2.5;t=1 \approx 1.08)$. Moreover, the correlations decrease for next-to-nearest and next-to-next-nearest neighboring filaments, since the intensity of the bright regions decreases as one moves away from the diagonal in Figs. \ref{fig:2050}($b_1$)-($b_3$), e.g. compare $g^{(2)}_{1,-1}(x=-1.7,x'=2.5;t=1) \approx 1.08$ to $g^{(2)}_{1,-1}(x=-3.7,x'=2.5;t=1) \approx 1.04$. 
We also remark that the correlation structures between the remaining components exhibit similar to the above-described characteristics (not shown here). 

\section{Conclusions and Outlook}
\label{conclusion} 

We have explored the many-body effects on the ground state properties
(and associated transitions) of a harmonically trapped spin-1 Bose gas upon varying the linear and quadratic Zeeman energy parameters as well as its breathing dynamics induced by quenching the external trapping frequency. 
To capture the different phases associated with second-order quantum phase transitions in the ground state of the system we resort to the population imbalance between the components of the spinor gas. 
We reveal how the boundaries of these phases are altered in the presence of intra- and inter-component correlations and for different particle numbers. 
On the other hand, the time-evolution of the one- and two-body density distributions of the individual components enable us to characterize the quench-induced breathing dynamics of our system. 
In this way the imprint of correlations in the course of the evolution is analyzed for different values of the spin-dependent interactions, and thus residing in distinct phases, spin-independent coupling constants and particle numbers. 

Focusing on the different magnetic phases emanating in the ground state of our system, we find that the inclusion of correlations is negligible concerning first-order transitions while the boundaries of the second-order ones are significantly altered. 
We unveil that for both ferromagnetic and anti-ferromagnetic spin-spin interactions, the involved transition borders are shifted leading to a decreased interval in terms of the linear and quadratic Zeeman parameters where the AF and BA phases are entered. 
Note that the aforementioned phases are characterized by a superposition of spin-states, thus demonstrating that correlated systems favor ground states where the bosons become polarized in a single spin-component. 
Additionally, by comparing the phase boundaries corresponding to second-order transitions for a varying particle number, it is showcased that the correlation effects on the emergent phase diagram become more prominent in the few-body scenario (i.e., $N=5-20$). 
Furthermore, by considering an adequately large number of bosons and operating within the MF realm we are able to recover the theoretical predictions in the thermodynamic limit associated with the absence of interparticle correlations. 
Already for $N=50$ the two descriptions become quite proximal and by $N=500$, the large particle limit is reached.

Turning to the dynamical response of the spinor gas it is shown that its breathing frequency is independent of the spin-state and the initial magnetic phase. 
We illustrated that the presence of intra- and intercomponent correlations leads to the formation of filamentary patterns in the one-body density of each participating component. 
The number of these filaments is found to increase for larger spin-independent interaction strengths or a smaller number of particles while keeping fixed all other system parameters. 
Interestingly, we demonstrate that for an initial BA state a spin-flip dynamics takes place, coherently transferring bosons from the $\alpha=\pm 1$ components to the $\alpha=0$ one and vice versa in the course of the evolution. 
We also showcase that this intercomponent particle transfer process can be controlled by means of adjusting either the strength of the linear Zeeman term or the quench amplitude. 

To further expose the effect of correlations during the dynamics we inspect the coherence losses and the two-body correlations within and between the spin components. 
It is shown that coherence is maintained within each filament during the dynamics, while significant losses of
coherence occur between the nearest and next-to-nearest neighboring filaments. 
Most importantly, we find that irrespectively of the spin-dependent and spin-independent interaction strength as well as the  number of bosons, two particles within a filament exhibit an anti-correlated character while particles between neighboring filaments are correlated to each other throughout the evolution. 

There is a variety of possible fruitful directions that can be pursued in future investigations. 
A straightforward extension of the present work would be to examine the correlation effects in the phase diagram of 
spinor Bose gases with higher than unity total spin \cite{schmaljohann2004dynamics}. 
In this way, it would be also possible to infer the dependence of correlations with respect to 
the total spin of the system. 
Another fruitful prospect is to unravel and possibly control the spin dynamics of a spin-1 Bose gas upon its exposure 
to a spatially dependent external magnetic field \cite{koutentakis2019probing}.
In addition, the possibility of quenches in the linear and quadratic Zeeman energy
shift parameters in order to examine abrupt transitions between different
phases promises to offer some interesting dynamics and possible metastable
states.
It is also worthwhile to examine the many-body dynamics of non-linear excitations in the form of dark-dark-bright or dark-bright-bright 
solitonic entities \cite{zhang2007solitons,ieda2004exact,romero2019controlled} in order to inspect the potential presence of a decay mechanism or other peculiar quantum properties already known to emerge for less complex soliton structures \cite{katsimiga2018many,katsimiga2017many,katsimiga2017dark}. 
Moreover, recent experiments have utilized controlled spin mixing interaction dynamics and generated fully entangled $^{87}$Rb spinor condensates, for instance the twin-Fock state~\cite{liyou1}, i.e., with exactly half of the atoms each in the spin components $\alpha=\pm 1$, and the analogous balanced spin-1 Dicke state~\cite{liyou2} involving all three spin ($\alpha=0$ and $\alpha=\pm 1$) components. It would be especially interesting to examine, within the ML-MCTDHX framework, the (less than $1$\%) deviations from nearly perfect coherence in such settings. Naturally, all of the present considerations have been constrained to the one-dimensional case. Thus, it is also of particular interest to generalize relevant considerations to higher dimensions where topologically charged configurations can arise (including especially complex ones such as skyrmions, monopoles, knots, etc.).

\begin{acknowledgments}
P.S. gratefully acknowledge financial support by the Deutsche Forschungsgemeinschaft (DFG) in the framework 
of the SFB 925 ``Light induced dynamics and control of correlated quantum systems''. 
S. I. M. acknowledges financial support in the framework of the Lenz-Ising Award of the University of  Hamburg. 
The authors are thankful to G. M. Koutentakis for useful discussions and a careful reading of the manuscript. 
This material is based upon work supported by the US National Science
Foundation under Grants No. PHY-1602994 and DMS-1809074
(PGK). PGK also acknowledges support from the Leverhulme Trust via a
Visiting Fellowship and thanks the Mathematical Institute of the University
of Oxford for its hospitality during part of this work.
\end{acknowledgments}

\bibliography{references}

\end{document}